\documentclass[aip, jcp,11pt,reprint]{revtex4-1}

\usepackage{graphicx}   
\usepackage{dcolumn}    
\usepackage{bm}         
\usepackage{amsmath}         
\usepackage[T1]{fontenc}       
\usepackage{color,soul}
\usepackage{multirow}

\newcommand{\eqn}[1]{Eq.~(\ref{#1})}

\newcommand{\bra}[1]{\big< \, #1 \, \big|}
\newcommand{\ket}[1]{\big| \, #1 \, \big>}
\usepackage{bm}         
\usepackage{color,soul}

\begin{document}

\title{Vibrational tunneling spectra of molecules with asymmetric wells:
a combined vibrational configuration interaction and instanton approach}


\author{Mihael Erakovi\'{c}}
\affiliation{Department of Physical Chemistry, Ru{\dj}er Bo\v{s}kovi\'{c} Institute,
Bijeni\v{c}ka Cesta 54, 10000 Zagreb, Croatia}

\author{Marko T.~Cvita{\v s}}
\email{mcvitas@phy.hr}
\affiliation{Department of Physics, Faculty of Science, University of Zagreb,
Bijeni\v{c}ka Cesta 32, 10000 Zagreb, Croatia}

\date{\today}

\begin{abstract}
A combined approach that uses the vibrational configuration interaction (VCI) and
semiclassical instanton theory was developed to study vibrational tunneling
spectra of molecules with multiple wells in full dimensionality.
The method can be applied to calculate low-lying vibrational states in the systems
with arbitrary number of minima, which are not necessarily equal in energy or shape.
It was tested on a two-dimensional double-well model system and on malonaldehyde
and the calculations reproduced the exact quantum-mechanical results with
high accuracy. The method was subsequently applied to calculate vibrational
spectrum of the asymmetrically deuterated malonaldehyde with non-degenerate
vibrational frequencies in the two wells.
The spectrum is obtained at a cost of single-well VCI calculations used to
calculate the local energies. The interactions between states of different
wells are computed semiclassically using instanton theory at a comparatively
negligible computational cost.
The method is particularly suited to systems in which the wells are separated
by large potential barriers and tunneling splittings are small, e.g,
in some water clusters, when the exact quantum-mechanical methods come at
a prohibitive computational cost.
\end{abstract}

\maketitle

\section{Introduction}
Physical systems with multiple energetically stable minima are ubiquitous in
chemistry and physics \cite{BellBook}. Bound states that are localized in
such wells, separated by potential barriers, interact via quantum tunneling,
which results in observable shifts of their energies
\cite{Hund1927tunnel,Benderskii}.
For equivalent, symmetry-related wells, the states that would be degenerate
in the absence of tunneling, produce a splitting pattern of energy levels.

Molecules and molecular complexes with two or more equivalent
stable configurations are multidimensional systems that display these
effects in their vibrational spectrum.
The inversion of ammonia \cite{Urban1981},
proton tunneling in malonaldehyde \cite{Firth1991malonaldehyde}, 
double proton transfer in porphycene \cite{Mengesha2013}
or bond rotation in vinyl radical \cite{Tanaka2004}
are examples of symmetric double-well systems that produce
measurable tunneling splittings (TS) of their vibrational state energies.
Water clusters are prototype multiwell systems that exhibit nontrivial
splitting patterns caused by tunneling rearrangements between many
stable configurations of the cluster \cite{Waterclust}.

The asymmetric systems, which have non-equivalent wells, have been less
studied. When the state energies of different wells are in resonance,
the tunneling dynamics will again cause the delocalization of
the wavefunction across the wells and the energy shifts in the spectrum
\cite{Benderskii}. Away from the resonance, the states remain localized
in one well. The asymmetry can be induced in symmetric molecular systems
by asymmetric isotopic substitutions \cite{Jahr2020}. The normal modes and
vibrational frequencies in equivalent symmetry-related potential wells
then differ and the correspondence of the vibrational wavefunctions of
different wells is not preserved in general.
As an example, the malonaldehyde molecule deuterated at D7/D9 position
(see Figure \ref{normal_modes}) thus has an asymmetric level structure with the localized
states and those that are delocalized across the multiple minima \cite{Jahr2020}.
A mixing angle between the left-right ground vibrational states of partially
deuterated malonaldehyde has been determined
experimentally \cite{Baughcum1981malonaldehyde}.
Further examples of the mixing have been studied in HF$-$HD dimer
\cite{Zhang1995} and partially deuterated vinyl radical \cite{Smydke2019},
CHD$-$CH, using full-dimensional exact calculations.
The splitting pattern in partially deuterated water trimers
HDO(H$_2$O)$_2$ and D$_2$O(H$_2$O)$_2$ have been determined in experiment
\cite{Liu1996trimer} and by us using instanton theory \cite{Erakovic2021}.

\begin{figure}
\begin{center}
{\includegraphics[width=8cm]{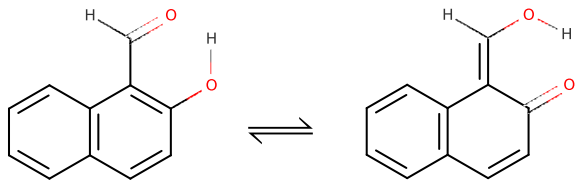}}
\caption{Tunneling tautomers of 2-hydroxy-1-naphthaldehyde.}
\label{naphtaldehyde}
\end{center}
\end{figure}

The asymmetry in molecules can also be found in some tautomers.
In this case, potential energy surface (PES) does not possess a symmetry
relating the wells and their shapes, and the minimum energies
are different. A possible candidate belonging to this class is
2-hydroxy-1-naphthaldehyde, shown in Figure \ref{naphtaldehyde}.
Hydroxyl proton forms a hydrogen bond with the oxygen atom of
the carbonyl group, and can tunnel to it to form a tautomer,
which is a local minimum.

\begin{figure}
\begin{center}
{\includegraphics[width=8cm]{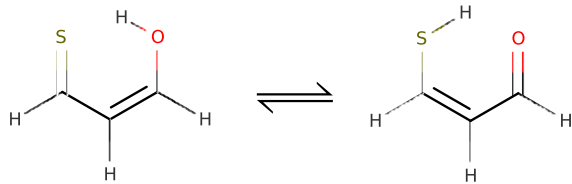}}
\caption{Tunneling tautomers of thiomalonaldehyde.}
\label{thiomalonaldehyde}
\end{center}
\end{figure}

Thiomalonaldeyde has two nearly degenerate minima in the form of enol
and enethiol tautomers, shown in Figure \ref{thiomalonaldehyde}.
Enethiol is about 70 cm$^{-1}$ more stable \cite{Gonzalez1997},
with the barrier height
to interconversion slightly lower than in the malonaldehyde.
This implies that the TS is similar in magnitude
to the energy asymmetry of the wells and the states in different
wells that lie below the barrier are expected to interact.
Interestingly, it has been suggested \cite{Gonzalez1997} that
the replacement of hydrogen, shared by the hydrogen bonds
OH$-$S and SH$-$O, by deuterium reverses the stability order
of tautomers due to zero-point energy effect.

The asymmetry can also be caused by environment. Molecules in
rare gas matrices can have energy asymmetry between the wells
comparable to their TS in isolation.
Delocalization of the tunneling hydrogen was observed
\cite{Bondybey1984}
in 9-hydroxyphenalone embedded in a neon matrix. Molecules in crystals
in the vicinity of a suitable guest molecule can also have
comparable energies of the splitting and energy asymmetry
of the wells \cite{Oppenlander1989}.

Quantum tunneling has also been observed in macroscopic systems.
Tunneling of Bose-Einstein condensates \cite{Hall2007},
electron spin tunneling in  the nanomagnetic molecules
\cite{Takahashi2011} or the tunneling of magnetic flux in
superconducting circuits based on Josephson junctions
\cite{Johnson2005} are some recent examples. In a collective
macroscopic variable, these processes can be described by
a double well with externally controllable parameters that
can induce asymmetry between the wells.

Calculation of TS in moderately large molecules is
prohibitively costly. Exact variational methods for determining the
bound states of molecules scale exponentially with the basis set size
while large basis sets are often required \cite{Felker2019}.
Basis functions need to span over two or more wells sufficiently
densely to obtain enough resolution to extract
the splittings from the difference of the energies in their spectrum.
The asymmetry of the wells also suggests that the symmetry cannot
be used to reduce the size of the problem. Full-dimensional studies of
malonaldehyde using multiconfigurational
time-depedent Hartree \cite{Schroder2014,Hammer2012} (MCTDH),
variational calculations on HF dimer \cite{Felker2019} or
H$_2$O dimer \cite{Leforestier2012} represent the state-of-the-art
calculations of the vibrational levels using formally exact methods.

A direct calculation of TS in larger systems can be performed using
a recently developed path integral molecular dynamics method
\cite{Vaillant2018dimer} based on the potential sampling around
the minimum action paths (MAP) connecting different wells.
The multi-well splitting patterns of water trimer and hexamer
\cite{Vaillant2019pimd} were obtained in this way using a matrix model
of hamiltonian in the basis of local vibrational states.
The tunneling matrix (TM) elements are extracted from the zero-temperature
limit of the partition function, which means that the method only works
for the vibrational ground-state in symmetric well systems.

Alternatively, the TM elements can be estimated using
semiclassical methods. From that class, the instanton method, which
comes in several forms
\cite{Benderskii1997,Smedarchina2012rainbow,tunnel,Milnikov2001},
has some particularly appealing features.
It can be applied in Cartesian coordinates \cite{tunnel,Erakovic2020}
to any molecule without modification.
Numerically, it relies on the optimization of the minimum action path
(MAP) that connects the symmetry-related minima \cite{Cvitas2016instanton},
and requires the potential and hessians of the potential along the MAP
to evaluate the splittings. It thus relies on a modest number of potential
and gradient evaluations \cite{Cvitas2018instanton} in comparison with
the exact quantum-mechanical (QM) methods.
This allows one to perform calculations in full dimensionality or
in combination with on-the-fly evaluation of the electronic potential.
Additionally, its accuracy is higher for large barriers and small splittings.
Precisely in this regime, the exact variational methods become inefficient
and resource intensive.

The first derivation of multidimensional instanton theory was accomplished
by means of Jacobi fields integration (JFI) \cite{Milnikov2001}.
JFI method has been used to determine TSs for a range of symmetric
double-well systems, such as malonaldehyde
\cite{Milnikov2001,Cvitas2016instanton,Cvitas2018instanton},
vinyl radical \cite{Milnikov2006},
and formic acid dimer \cite{Milnikov2005formic}.
The instanton method was later rederived in the ring polymer form (RPI)
\cite{tunnel}, which could treat asymmetric potentials along MAPs and
multiple wells. The RPI was used to calculate and interpret experimental
ground-state splitting patterns of water clusters in terms of their
rearrangement dynamics \cite{Waterclust} for the dimer,
\cite{tunnel,Vaillant2018rotation,Zwart1991dimer} trimer
\cite{tunnel,Keutsch2003review}
hexamer \cite{hexamerprism} and octamer \cite{octamer}.
We recently generalized the JFI method \cite{Erakovic2020} to treat
the multi-well systems and used it to explain the ground-state
splitting pattern of 320 states in the water pentamer in terms of
five dominant rearrangement pathways \cite{Cvitas2019}.
The extension of the method to low-lying vibrational states
\cite{Erakovic2020vib} is based on work of Mil'nikov and Nakamura
\cite{Milnikov2005} and forms the groundwork of calculating
the TM elements between local vibrational
states of different wells in the present study below.

Weakly biased double-well systems have been considered in previous work
by several authors. Analytical results in one dimension have been obtained
using semiclassical WKB and instanton methods.
Garg has demonstrated \cite{Garg2000} that the instanton and the WKB method
with Herring formula \cite{Herring1962} give equivalent results for
TS in symmetric systems.
Cesi et al.~\cite{Cesi1991} considered a one-dimensional (1D) double-well
with the shape asymmetry and no energy asymmetry using instantons and
obtained an expression for the ground-state TS.
An approximate solution for a 1D double well with a weak bias
was also obtained by Mugnai and Ranfagni \cite{Mugnai1985}, using instantons
based on the MAP that does not fully connect the minima of the two wells.
Leggett {\em et al} obtained a solution \cite{Leggett1987} by
adding a parabolic correction potential to remove the asymmetry between
the wells, the contribution of which was then subsequently subtracted from
the action integral. Dekker \cite{Dekker1987} derived the ground-state
TS from the quantization condition by asymptotic matching of
the semiclassical wavefunction in the barrier to the parabolic cylinder
wavefunctions of harmonic oscillators in the two wells.
Song \cite{Song2008,Song2015} extended Dekker's method \cite{Dekker1987}
(as have Halataie and Leggett \cite{Leggett_prep} done independently)
to obtain the TS in vibrationally excited states of asymmetric 1D potentials
with arbitrarily large shape and energy asymmetry.
Song also showed \cite{Song2015} that the instanton
wavefunctions with the Herring formula in a $2\times2$ matrix model
give equivalent results to those obtained by Dekker's method \cite{Dekker1987}.

In multidimensional systems, tunneling can be assisted or supressed by the
excitation of transversal vibrational modes \cite{Milnikov2005,Siebrand2013}.
In the presence of asymmetry, the excited states of one well can be
in a resonance with the states of another well with a different set of
local quantum numbers, which results in a delocalization of
the wavefunction across these wells \cite{Song2015}.
Benderskii {\em et al} devised a multidimensional perturbative instanton
method \cite{Benderskii1999}
in which they treat the asymmetry of the potential in an analytic two-dimensional
model as a correction of first order in $\hbar$, same as energy. In this way,
the MAP remains symmetric and the asymmetry is moved to the transport equation
along with energy.
They also show that the equivalent expressions for the TS are
obtained using the instanton quantization condition of Dekker \cite{Dekker1987}
and using the instanton or WKB wavefunctions with Herring formula
\cite{Herring1962} in 1D. The method was applied to calculate TSs in
excited vibrational states of malonaldehyde \cite{Benderskii2000}
with the asymmetric isotopic substitutions using a fit of model
potential parameters to quantum-chemical data.
The extensions of the RPI and JFI method to the ground-states of
the asymmetric systems with a weak bias have recently been derived
and applied to partially deuterated malonaldehyde \cite{Jahr2020}
and water trimer \cite{Erakovic2021}, respectively.

The object of this paper is to propose a method for calculating
vibrational tunneling spectrum of multi-well systems of mid-sized
molecules that are outside reach of the exact quantum methods.
For this purpose, we extend the usual $2\times2$ matrix model to
the $\sum_m N_m \times \sum_m N_m$ model, which represents
the molecular Hamiltonian in the basis of all $N_m$ local vibrational
states of each well $m$. We rederive
a generalized Herring formula \cite{Herring1962,Benderskii1999}
in order to calculate the off-diagonal TM elements
that represent the interaction of local vibrational states of
different wells. The semiclassical wavefunctions at the dividing plane,
in the barrier that separates the wells, are obtained using
the recently generalized JFI method \cite{Milnikov2005,Erakovic2020}.
The JFI wavefunctions are thus used to calculate the couplings between
states that have different energies and normal-mode excitations
for the first time.
The diagonal energies of the local vibrational states can be calculated
using any accurate quantum method with a basis set that spans only one
well. Vibrational configuration interaction
\cite{Carter1998,Christoffel1982,Bowman1979}
(VCI) is used in this work. The effect of rotations is neglected.

The method, presented in Section II, allows one to study vibrational
structure in asymmetric systems with multiple wells, separated by
large potential barriers, in an approximate manner.
The accuracy of the method is tested on a two-dimensional
double-well model in Section III~A. In Section III~B, it is applied to
the (symmetric) malonaldehyde molecule, which tests the accuracy
of the matrix model using a combination of VCI and JFI matrix elements
on a realistic PES, in vibrationally excited states, against 
the exact MCTDH calculations. Vibrational tunneling spectrum
of the partially deuterated malonaldehyde is calculated in
Section III~C, which features  the mixing of inequivalent well
states due to tunneling. The paper concludes in Section IV.

\section{TUNNELING MATRIX}
Without the loss of generality, we start by considering a system with
{\em two} minima separated by a large potential barrier. The minima, denoted
as `left' (L) and `right' (R), are not necessarily symmetric either
in shape or energy. For low-energy spectra, the vibrational Hamiltonian can be 
represented in the basis of states that are localized in the wells,
$\{ \phi_i^{(\rm L)}, \phi_j^{(\rm R)} \}$, as
\begin{equation}
    \begin{pmatrix}
    \mathbf{H}^{(\rm L)} & \mathbf{h} \\
    \mathbf{h}^{\top} & \mathbf{H}^{(\rm R)} \\
    \end{pmatrix}.
\label{model_hamiltonian}
\end{equation}
Square blocks $\mathbf{H}^{(\rm L / R)}$ are formed using basis functions
of the same minimum and are not necessarily of equal size.
Their off-diagonal elements describe the interaction between different
basis functions localized in the same minimum and can be made small by
a suitable choice of the basis.
In the instanton theory of tunneling splittings, the usual presumption is
that the local vibrational wavefunctions are harmonic oscillator states.
In that case, the off-diagonal terms describe anharmonic contributions that
originate from the difference between the actual and the harmonic potential.

In our approach here, we replace the harmonic surface of each well by
an $n$-mode representation \cite{Bowman2003,Rauhut2004} 
of the well potential and calculate local
eigenfunctions and eigenvalues using vibrational self consisten field
(VSCF) and vibrational configuration interaction (VCI) methods
\cite{Carter1998,Christoffel1982,Bowman1979}.
The technical details of the calculations are described in Appendix C.
Using the more accurate local wavefunctions as a basis reduces the magnitude
of the off-diagonal matrix elements in $\mathbf{H}^{(\rm L / R)}$, which
we then neglect. The matrices $\mathbf{H}^{(\rm L / R)}$ become diagonal
and the diagonal matrix elements are referred to as the local
vibrational energies of the left/right (L/R) well. For symmetric wells,
the local energies are doubly degenerate.

The block $\mathbf{h}$ in matrix (\ref{model_hamiltonian}) contains
the TM elements that describe the interaction of local
wavefunctions of the left and right minimum. The exact quantal calculation
of these elements requires a large basis set that can accurately represent
the form of the wavefunction inside the barrier. Instead, we obtain them
by means of Herring formula \cite{Herring1962} in combination with
the semiclassical wavefunctions from the instanton theory
\cite{Milnikov2005,Erakovic2020vib}. Since the only effect of matrix
(\ref{model_hamiltonian}) is to mix local wavefunctions of different
minima via tunneling, we refer to it as the tunneling matrix \cite{tunnel}.

We now derive the Herring formula without the usual assumptions of
the two-state model and the L/R symmetry. Rather,
we consider Schr\"{o}dinger equation with Hamiltonian (or tunneling) matrix
(\ref{model_hamiltonian}) from which it follows that
\begin{align}
    \nonumber
    \hat{H}\phi_i^{(\rm L)} &= E_i^{(\rm L)}\phi_i^{(\rm L)}+h_{ik}\phi_k^{(\rm R)}, \\
    \hat{H}\phi_j^{(\rm R)} &= E_j^{(\rm R)}\phi_j^{(\rm R)}+h_{kj}\phi_k^{(\rm L)},
\label{hamiltonian_on_localized_states}
\end{align}
where $E_i^{(\rm L/R)}$ are the local vibrational energies and the summation over
repeated indices is assumed.
Next, a dividing plane is defined inside the barrier via the implicit equation
$f_{\rm D}(\mathbf{x})=0$, which separates the left from the right minimum.
Eqns.~(\ref{hamiltonian_on_localized_states}) are multiplied by $\phi_j^{(\rm R)}$ and
$\phi_i^{(\rm L)}$, respectively, subtracted and integrated over the left part
of the domain (i.e., over the space on the `left' side of the dividing plane).
The local wavefunctions $\phi_i^{(\rm L/R)}$, either harmonic or VCI, have been
obtained as eigenfunctions of a hermitian matrix and are therefore taken to be
orthonormal.
For a sufficiently high barrier, the wavefunctions $\phi_i^{(\rm L/R)}$ can be
considered small in the R/L domain, respectively. We thus neglect the integrals
involving the like products $\phi_i^{(\rm R)}\phi_i^{(\rm R)}$ in the L volume
and extend the integrals involving $\phi_i^{(\rm L)}\phi_j^{(\rm L)}$ over the entire domain
to produce $\delta_{ij}$.
The integrals involving the mixed products $\phi_i^{(\rm L)}\phi_j^{(\rm R)}$ have also been
neglected. The error introduced by the neglect of these terms outside the resonance, i.e.,
for $ E_i^{(\rm L)} \ne E_j^{(\rm R)}$, is analysed in Appendix B.
The TM element is then expressed as
\begin{align}
\nonumber
h_{ij} &=\int_{\rm L} \left( \phi_i^{(\rm L)} \hat{H}\phi_j^{(\rm R)} -
\phi_j^{(\rm R)} \hat{H}\phi_i^{(\rm L)} \right) d\mathbf{x} \\
\nonumber
&=\frac{1}{2} \int_{\rm L} \nabla \left( \phi_j^{(\rm R)} \nabla \phi_i^{(\rm L)} -
  \phi_i^{(\rm L)} \nabla \phi_j^{(\rm R)} \right) d\mathbf{x} \\
&=\frac{1}{2} \int \left( \phi_j^{(\rm R)} \frac{\partial}{\partial S} \phi_i^{(\rm L)} -
\phi_i^{(\rm L)} \frac{\partial}{\partial S} \phi_j^{(\rm R)} \right)
\delta(f_{\rm D}(\mathbf{x})) d\mathbf{x},
\label{herring_derivation}
\end{align}
where, in the last step, we use the divergence theorem to turn the spatial integration
into the integral over the dividing plane. $S$ in \eqn{herring_derivation} denotes
the coordinate that describes an orthogonal shift from the dividing plane.

Local wavefunctions, that we designed to calculate the local vibrational energies
on the diagonal of matrix (\ref{model_hamiltonian}), are constructed using VSCF/VCI on
an approximate PES (see Appendix C) and their accuracy drops inside the barrier that
separates the wells.
In order to evaluate the surface integral in Herring formula, \eqn{herring_derivation}, inside
the barrier, we employ the JFI wavefunctions instead, which we recently derived
in Ref.~\onlinecite{Erakovic2020vib}.
These are based on the WKB method in which the energy is treated as a term of order $\hbar^1$
and is moved to the transport equation, leaving the Hamilton-Jacobi equation energy independent.
It was shown that this approach gives equivalent results to the standard WKB method
in 1D \cite{Garg2000}. Moreover, the ground-state TS obtained from Herring
formula using the ground-state JFI wavefunctions \cite{Erakovic2020vib} is identical to the
standard instanton result derived from the steepest descent approximation of the partition function
in the path integral formulation \cite{Erakovic2020}.

The characteristic of the Hamilton-Jacobi equation that connects the minimum of
a well to a point in configuration space obeys the equation \cite{Erakovic2020vib}
\begin{equation}
    \frac{d^2}{d \tau^2} {\mathbf x} = \nabla V,
\end{equation}
and represents a classical trajectory ${\mathbf x}(\tau)$ on the inverted PES,
parametrized by the `imaginary' time $\tau$. In order to represent the quantities
in the neighborhood of the characteristic, $N$ local coordinates
$(S,\Delta \mathbf{x})$ are defined \cite{Milnikov2005,Erakovic2020vib},
where $S$ is the mass-scaled arc length distance from the minimum along
the characteristic and $\Delta {\mathbf x}$ is the orthogonal shift from
the nearest point on the characteristic. The classical momentum is defined as
\begin{equation}
p_0^{(\rm L/R)}=\frac{dS}{d\tau}=\sqrt{2(V-V_{\rm min}^{(\rm L/R)})},
\label{momentum}
\end{equation}
and $S$ can be used, instead of $\tau$, to reparametrize the characteristic.
Local wavefunctions in the harmonic vicinity of the characteristic are
obtained by integrating the Hamilton-Jacobi and transport equations
on the characteristic \cite{Erakovic2020vib} as
\begin{align}
    \nonumber
    \phi_{\nu}^{(\rm L/R)} = &\sqrt[4]{\frac{\det{{\mathbf A}_0^{(\rm L/R)}}}{\pi^N}}
    \sqrt{\frac{\left(2\omega_{\rm e}^{(\rm L/R)}\right)^{\nu}}{(2\nu-1)!!}} \\
\nonumber
& \times  \left(F^{(\rm L/R)}+{\mathbf U}^{(\rm L/R)\, \top}\Delta{\mathbf x}\right)^{\nu} 
    {\rm e}^{-\int_0^{S} p_0^{(\rm L/R)}(S') dS'} \\ 
&  \times {\rm e}^{- \frac{1}{2}\int_0^{S}
\frac{\rm{Tr} \left({\mathbf A}^{(L/R)}-{\mathbf A}_0^{(L/R)}\right)}{p_0^{(\rm L/R)}(S')} dS'
-\frac{1}{2}\Delta{\mathbf x}^{\top} {\mathbf A}^{(\rm L/R)}\Delta {\mathbf x}}.
\label{localized_wavefunctions}
\end{align}
For vibrationally excited states, the label $\nu$ in \eqn{localized_wavefunctions}
is the number of quanta in the excited vibrational mode of frequency $\omega_{\rm e}$.
Matrices ${\mathbf A}^{(\rm L/R)}$ are Gaussian widths of the wavefunction in
the directions orthogonal to the characteristic and are obtained from
\begin{equation}
p_0^{(\rm L/R)} \frac{d}{d S} {\mathbf A}^{(\rm L/R)} =
{\mathbf H}(S)-\left( {\mathbf A}^{(\rm L/R)} \right)^2.
\label{riccati}
\end{equation}
${\mathbf H}(S)$ in \eqn{riccati} is Hessian of the potential at $S$, which is used to 
approximate the potential up to quadratic terms in the neighborhood of the characteristic.
The initial condition for \eqn{riccati} at the minimum is
${\mathbf A}_0^{(\rm L/R)} = \left({\mathbf H}_0^{(\rm L/R)}\right)^{{1}/{2}}$,
where ${\mathbf H}_0^{(\rm L/R)}$ is Hessian at the L/R minimum.
For vibrationally excited states, the prefactor in the parenthesis in \eqn{localized_wavefunctions}
contains terms $F(S)$ and $\mathbf{U}(S)$, which are defined via equations
\begin{align}
    \nonumber
    p_0^{(\rm L/R)} \frac{d}{dS} F^{(\rm L/R)} &= \omega_{\rm e}^{(\rm L/R)} F^{(\rm L/R)}, \\
    p_0^{(\rm L/R)} \frac{d}{dS} {\mathbf U}^{(\rm L/R)} 
    &= \omega_{\rm e}^{(\rm L/R)} {\mathbf U}^{(\rm L/R)} - {\mathbf A}^{(\rm L/R)} {\mathbf U}^{(\rm L/R)}.
\label{f_and_u_equations}
\end{align}
$F^{(\rm L/R)}$ terms account for the change in the amplitude of the excited-state wavefunction
along the characteristic, while the ${\mathbf U}^{(\rm L/R)}$ term describes the nodal plane.
The initial condition for $F^{(\rm L/R)}$ is found by matching the instanton wavefunction to
that of the harmonic oscillator at a small distance $S=\varepsilon$ from the minimum, as
$F^{(\rm L/R)}(\varepsilon) = {\mathbf U}_0^{(\rm L/R), \top} ({\mathbf x}(\varepsilon)-{\mathbf x}_{\rm min}^{(\rm L/R)})$.
The ${\mathbf U}_0^{(\rm L/R)}$ is the excited-state normal mode, i.e.,
the eigenvector of ${\mathbf H}_0$ having frequency $\omega_{\rm e}$, and serves
as the initial condition for ${\mathbf U}$ in \eqn{f_and_u_equations}.

The local instanton wavefunctions, \eqn{localized_wavefunctions}, for
the left and right minimum are next inserted into Herring formula
\eqn{herring_derivation}, without the previous assumption \cite{Erakovic2020vib}
that $\phi^{\rm (L)}$ and $\phi^{\rm (R)}$ refer to the excitation of the same
normal mode and the same number of quanta $\nu$. For that purpose, a connection
point $\mathbf{x}(S_{\rm cp})$ is chosen on the dividing surface
$f_{\rm D}(\mathbf{x})$ inside the barrier and characteristics determined,
which connect it to the minima on both sides of the dividing surface.
The shape of the characteristic between two points in configuration space
is determined by minimizing the Jacobi action \cite{Cvitas2016instanton}.
The surface integral in \eqn{herring_derivation} can then be computed
analytically \cite{Erakovic2020vib}.

This approach yields best results if the connection point is chosen so that both
wavefunctions are near their maxima in the dividing plane at the connection
point. This can be obtained by minimization of the sum of action integrals
$\int_0^{S_{\rm cp}^{(\rm L)}} p_0^{(\rm L)}(S') dS'+\int_0^{S_{\rm cp}^{(\rm R)}} p_0^{(\rm R)}(S') dS'$. 
For minima of the same energy, this procedure yields the minimum
action path (MAP) that connects the minima and any point on that path is
a suitable candidate for the connection point. The dividing surface can then be
chosen as the plane orthogonal to the MAP at the connection point.
If the minima do not have the same energies, but differ by the amount $d$,
this procedure is equivalent to determining the MAP
on the modified PES $\tilde{V}({\mathbf x}) = V({\mathbf x}) - \Theta (S-S_{\rm cp})d$, 
where $S_{\rm cp}$ is the position of the connection point on the characteristic
and $\Theta$ is the Heaviside step function.
In this case, the position of the connection point has to be given a priori,
and the resulting path will depend on its position.
The safest choice is to pick the connection point in the middle of the MAP,
which is expected to be near the maximum of the potential energy barrier.
For minima at different energies, the resulting MAP is going to have a
tangent dicontinuity at the connection point as $p_0^{\rm (L)} \ne p_0^{\rm (R)}$
at $S_{\rm cp}$. The tangent direction at the connection point is then
defined as the average tangent of its L and R limit at $S_{\rm cp}$.
Again, the dividing plane is taken to be orthogonal to the MAP
and the surface integral in \eqn{herring_derivation} is solved analytically.
The TM element then becomes
\begin{align}
    \nonumber
    h_{\nu \nu'} & = -\sqrt{\frac{\sqrt{\det' {\mathbf A}_0^{(\rm L)} \det' {\mathbf A}_0^{(\rm R)}}}{{\pi \det' \bar{{\mathbf A}}}}} \frac{p_0^{(\rm L)}+p_0^{(\rm R)}}{2} \\
    \nonumber
    &\left[\left( F^{(\rm L)} \right)^{\nu} \left( F^{(\rm R)} \right)^{\nu'} + \frac{1}{2}{\mathbf U}^{(\rm L)} \bar{\mathbf A}^{-1} {\mathbf U}^{(\rm R)} \delta_{1,\nu}\delta_{1,\nu'} \right] \\
    \nonumber
    &\sqrt{\frac{\left( 2\omega_{\rm e}^{(\rm L)} \right)^{\nu} \left( 2\omega_{\rm e}^{(\rm R)} \right)^{\nu'}}
     {(2\nu-1)!! (2\nu'-1)!!}}{\rm e}^{-\int_0^{S_{\rm cp}} p_0^{(\rm L)}d S -\int_{S_{\rm cp}}^{S_{\rm tot}} p_0^{(\rm R)}d S} \\
    &{\rm e}^{-\frac{1}{2}\int_0^{S_{\rm cp}} \frac{\rm{Tr} ({\mathbf A}^{(\rm L)} - {\mathbf A}_0^{(\rm L)})}{p_0^{(\rm L)}} dS -\frac{1}{2}\int_{S_{\rm cp}}^{S_{\rm tot}} \frac{\rm{Tr} ({\mathbf A}^{(\rm R)} - {\mathbf A}_0^{(\rm R)})}{p_0^{(\rm R)}} dS},
\label{h_ij}
\end{align}
with all quantities in the brackets evaluated at $S=S_{\rm cp}$. In \eqn{h_ij},
$S_{\rm tot}$ is the total length of the MAP,
$\bar{\mathbf A}=\left.\dfrac{{\mathbf A}^{(\rm L)}+{\mathbf A}^{(\rm R)}}{2}\right|_{\perp}$
and the symbol $\perp$ means that the tangent direction to the MAP was explicitly projected out.
det' in \eqn{h_ij} denotes the product of all non-zero eigenvalues. 
Matrices ${\mathbf A}_0$ have zero eigenvalues associated with the overal translations and rotations,
while $\bar{\mathbf A}$ has an additional zero eigenvalue associated with the tangent to the MAP.
For energy-equivalent minima, the tangent vector is an eigenvector of $\bar{\mathbf A}$ with
zero eigenvalue and the explicit projection to the orthogonal space is not needed.
The TM element in \eqn{h_ij} is valid for $\nu$, $\nu'=0-1$. For $\nu > 1$ and multiple excitations
in different modes, the TM element can still be evaluated using Herring formula and wavefunctions of
form (\ref{localized_wavefunctions}) using analytical integrals, but we have only implemented it
numerically, without trying to write down the explicit form. We also remark here that the wavefunction
in \eqn{localized_wavefunctions} for the multiply excited normal modes, $\nu>1$, does not correspond
to the harmonic oscillator wavefunction near the minimum, as the prefactor in
\eqn{localized_wavefunctions} is not a Hermite polynomial.
We further note that the TM element, (\ref{h_ij}), is not invariant with respect to the position
of the connection point $S_{\rm cp}$ unless the two local states are in resonance,
as shown in Appendix B.

\section{NUMERICAL TESTS}
Numerical tests were carried out on a model two-dimensional (2D) PES and on malonaldehyde
molecule with some atoms substituted with heavier isotopes.
MAPs that connect the minima were determined using the string method
\cite{Cvitas2016instanton,Cvitas2018instanton}.
The criterion for convergence was chosen to be the largest component of gradient of
Jacobi action perpendicular to the path and was set to $10^{-6} \ \rm{a.u.}$.
Number of beads used to discretize the string was 301 for model potential, which is much larger
than necessary for convergence, but was used to ensure that all results obtained using different
parameters of the potential are sufficiently converged. For potential with minima at different
energies, the dividing plane was set to pass through the central bead and perpendicular to the MAP.
In the case of malonaldehyde, the number of beads was 201 and the minima were oriented towards
the first neighboring bead in each step of the optimization to minimise the root-mean-square distance
between their geometries \cite{Cvitas2016instanton}. After optimization, Hessians of the potential were
determined at each bead on the MAP. Translations and rotations were explicitly projected
out from Hessians \cite{Erakovic2020}. Geometries along the path in mass-scaled Cartesian coordinates,
potential and Hessian matrix elements were parametrized by the arc length $S$ along the MAP
and interpolated using natural cubic splines.
Matrices $\mathbf A^{(\rm L / R)}$ in \eqn{riccati} were propagated using the previously described
approach \cite{Erakovic2020vib}, with the initial `jump' at $\varepsilon=0.1 \ \rm{a.u.}$ for
model potential and $\varepsilon = 0.25 \ \rm{a.u.}$ for malonaldehyde.
Fourth order Runge-Kutta method was used for integration of \eqn{riccati}.
Matrices ${\mathbf A}^{(\rm L / R)}(S)$ were saved at each bead and their matrix elements
interpolated using natural cubic splines, as for Hessians above. The interpolant was then used
to propagate $F^{(\rm L / R)}$ and $\mathbf U^{(\rm L / R)}$ in \eqn{f_and_u_equations} from
minima up to the the dividing plane.

The particular implementation of the VSCF/VCI method that is employed in our calculation here
is described in Appendix C. We determined the 1-mode and 2-mode terms of the PES and neglected
the terms beyond. In each normal mode,
the potential was evaluated at Gauss-Hermite discrete-variable-representation (DVR) points, which
correspond to the zeroes of Hermite polynomials. We used 8 DVR points for the 2D model potential
and 11 DVR points for malonaldehyde.
This approach utilizes the natural lengthscales of the harmonic oscillators in each normal
mode, which gives a balanced description of potential at different minima.
The 1-mode terms were then fitted to the eighth-order polynomials using linear regression.
For 2-mode terms, the potential was computed on a rectangular grid of DVR points determined above
and a fit was performed analogously. For each 1-mode potential, a quick QM calculation was performed
using sine DVR basis with 100 basis functions. The difference in the lowest two energies
from that calculation was used as a frequency for the harmonic oscillator basis set,
which was used to solve the VSCF equations. This approach provides a better basis for determining
the 1-mode potentials which quickly deviate from the harmonic curve, reducing the number of
basis functions needed to describe the 1-mode functions in VSCF.
$N_{\rm{basis}}=7$ and $N_{\rm{basis}}=16$ basis functions were used for each normal
mode for the 2D model potential and malonaldehyde, respectively, to converge the energies.
A larger basis should not be used, as functions corresponding to larger energies
penetrate into unphysical part of the fitted potential, which can cause appearance
of intruder states and worse energies. After VSCF calculation, the computed 1-mode functions
were used for VCISD calculation, where the highest excitation in each mode was limited to
6 in both, the 2D model system and malonaldehyde.

\subsection{2D MODEL POTENTIAL}
The 2D model potential with two minima, which we use in our test calculations below,
is defined by the following equations,
\begin{align}
\nonumber
&V({\mathbf x}) = \dfrac{\gamma_1 V^{(\rm L)} (\gamma_2 V^{(\rm R)}+d)}{\gamma_1 V^{(\rm L)}+\gamma_2 V^{(\rm R)}}, \\
\nonumber
&V^{(\rm L)}({\mathbf x}) = \frac{1}{2} \Delta \mathbf{x} ^{(\rm L) \: \top} {\mathbf U}_0^{(\rm L)}
\begin{pmatrix} \alpha_{1, \rm{L}} & 0 \\ 0 & \alpha_{2,\rm{L}} \end{pmatrix}
{\mathbf U}_0^{(\rm L) \: \top} \Delta \mathbf{x} ^{(\rm L)}, \\
\nonumber
&V^{(\rm R)}({\mathbf x}) = \frac{1}{2} \Delta \mathbf{x} ^{(\rm R) \: \top} {\mathbf U}_0^{(\rm R)}
\begin{pmatrix} \alpha_{1, \rm{R}} & 0 \\ 0 & \alpha_{2,\rm{R}} \end{pmatrix}
{\mathbf U}_0^{(\rm R) \: \top} \Delta \mathbf{x} ^{(\rm R)}, \\
\nonumber
&{\mathbf U}_0^{(\rm L)} = \begin{pmatrix} \cos{\theta} & -\sin{\theta} \\ \sin{\theta} & \cos{\theta} \end{pmatrix}, \qquad
{\mathbf U}_0^{(\rm R)} = \begin{pmatrix} -\cos{\theta} & \sin{\theta} \\ \sin{\theta} & \cos{\theta} \end{pmatrix}, \\
\nonumber
&\Delta \mathbf{x}^{(\rm L / R)} ={\mathbf x} - {\mathbf x}^{(\rm L)}, \qquad
{\mathbf x}^{(\rm L/R)} = \left( \pm \beta , 0 \right)^\top, \\
&\gamma_1=\frac{1+\frac{d^2}{V^{(\rm  L)}(\mathbf{x}^{(\rm R)})V^{(\rm R)}(\mathbf{x}^{(\rm L)})}}{1+\frac{d}{V^{(\rm R)}(\mathbf{x}^{(\rm L)})}}, \
\gamma_2=\frac{1+\frac{d^2}{V^{(\rm L)}(\mathbf{x}^{(\rm R)})V^{(\rm R)}(\mathbf{x}^{(\rm L)})}}{1-\frac{d}{V^{(\rm L)}(\mathbf{x}^{(\rm R)})}},
\label{sym_pot}
\end{align}
where ${\mathbf x}$ are not mass scaled. Minima are located at $\mathbf{x}^{(\rm L / R)}$.
Coefficients $\gamma_1$ and $\gamma_2$ are chosen so that in the vicinity of left minimum,
the potential is approximately harmonic and equals $V\approx V^{(\rm L)}$, while in
the vicinity of the right minimum, the potential is approximately harmonic and shifted
in energy by $d$, i.e., $V\approx V^{(\rm R)}+d$. $\alpha_{1,\rm L / R}$ and $\alpha_{2,\rm L / R}$
are eigenvalues of Hessian, while $\mathbf{U}_0^{(\rm L / R)}$ are normal modes. 
Parameter $\theta$ denotes the angle of inclination of normal mode to $x$ axis.
Mass of the  system was taken to be $m=3.5$ in both dimensions, so that the harmonic frequencies
are given by $\omega_{1/2}^{(\rm L / R)}=\sqrt{\alpha_{1/2, \rm L / R} / m}$. 

The above form of the potential can be used to independently vary harmonic frequencies $\omega_{1/2}^{(\rm R)}$,
by changing parameters $\alpha_{1/2, \rm R}$, or the shift $d$ without affecting the other parameters of either
the left or the right minimum. In this paper, the parameters of the left minimum were $\alpha_{1, \rm L}=1.6$,
$\alpha_{2, \rm L}=4.0$. The parameters of the right minimum were the same as the parameters of the left,
for the symmetric case with $d=0$. To obtain the asymmetric potentials below, one of the three 
parameters was varied, with $\alpha_{1, \rm R}$ going from $1.6$ to $36$, parameter $\alpha_{2, \rm R}$ 
going from $4$ to $49$ and $d$ going from $0$ to $1.1$. Positions of the minima were set with $\beta=2.0$ 
and the angle $\theta=\pi/12$. This angle corresponds to approximately equal 
contributions of $F^{(\rm L / R)}$ and $\mathbf{U}^{(\rm L / R)}$ in the TM elements
\cite{Erakovic2020vib}. Figure \ref{fig_1} shows the model potential for a selection of parameters
$\alpha_{1, \rm R}$, $\alpha_{2, \rm R}$ and $d$.

Frequency $\omega_1$ is the lower frequency and the MAP enters the minima along the corresponding normal mode.
Consequently, $\omega_1$ does not contribute towards the zero-point energy in the plane orthogonal to the MAP.
The effective barrier for the tunneling motion from the ground state in the left minimum, corrected by
the zero-point motion contribution, can be defined as
\begin{equation}
V^{(\rm L )}_{\rm eff} = V_{\rm max}
+\frac{1}{2}(\lambda^{(\rm L)}-\omega_1^{(\rm L )}-\omega_2^{(\rm L )}).
\label{effective_barrier_ground_state}
\end{equation}
$V_{\rm max}$ in \eqn{effective_barrier_ground_state} is the maximum of the potential
$V(S_{\rm max})$ along the MAP. $\lambda^{(\rm L )}$ is the non-zero eigenvalue of
$\mathbf A_{\perp}=\mathbf P \mathbf A \mathbf P$ matrix, where $\mathbf P$ projects out
the tangent direction to the MAP at $S=S_{\rm max}$. The effective barrier
can be defined for other states similarly.
Figure \ref{fig_2} (in the second column panels) shows that for the symmetric case,
$\omega_1^{(\rm R)}=\omega_1^{(\rm L)}$,
the JFI theory provides accurate tunneling splittings in the ground 
state and in the second excited state, which corresponds to the excitation of
the transversal frequency $\omega_2$.
In the first excited state, JFI theory slightly overestimates the TS.
In that state, the effective barrier is much smaller and equals $V_{\rm eff}=0.545$,
in contrast with the barriers of $1.221$ and $0.976$ for the ground and second excited state.
This overestimation is a known property of the instanton method
\cite{Erakovic2020vib}.
In the symmetric case, the only contribution to the splitting comes from
the off-diagonal matrix elements, so that the harmonic and VCI energies yield
same results. However, it can be observed
(from the first column panels in Figure \ref{fig_2})
that the harmonic vibrational energies overestimate the exact QM
energies by $3-5\% $. 

As the frequency $\omega_1^{(\rm R)}$ is increased and the difference in 
the local L/R energies begins to contribute to the overall splitting, the TSs computed
using harmonic energies quickly begin to deviate from the QM values due to the neglect
of anharmonicities, which no longer cancel out.
When the difference in the lower frequency, $\omega_1^{ (\rm R)}-\omega_1^{(\rm L)}$,
is only $0.02$, which corresponds to the asymmetry ($\Delta \omega_1 / \omega_1^{(\rm L)}$) of $3 \%$,
the error in the TS of the ground state is $24 \%$, whereas it is $90 \%$
for the transversal mode ($\omega_2$) excitation.
A larger error in the excited state reflects the fact that the local excited-state
wavefunction penetrates deeper into the barrier, where anharmonicity is larger.
However, the VCI energies correctly account for the anharmonicity
and provide an excellent agreement, as can be observed 
in Figure \ref{fig_2}, both in the absolute energies
(first column panels in Figure \ref{fig_2}) and in the TSs (second column panels).

With a further increase in the frequency $\omega_1^{(\rm R)}$, different local
vibrational states of the left and right minimum enter into resonance and
vibrational energies exhibit avoided crossings, shown in frames IV$-$VI
of the top panel in Figure \ref{fig_2}. Harmonic energies do not provide accurate
positions of these avoided crossings, as seen in Figure \ref{fig_2},
due to errors in the local energies. In the case of the avoided crossing between
the higher-frequency $\omega_2^{(\rm L)}$-excited state of the left minimum and
the $\omega_1^{(\rm R)}$-excited state of the right minimum,
shown in frame IV of the top panel in Figure \ref{fig_2},
the error in the position of the avoided crossing (in $\omega_1^{ (\rm R)}-\omega_1^{(\rm L)}$)
using harmonic energies is $16 \%$ (and falls outside the frame IV in Figure \ref{fig_2}).
VCI energies, shown magnified in Figure \ref{fig_3} together with the exact QM energies, provide
a significantly more accurate position with the error of only $0.4 \%$.
The small discrepancy can be attributed to the fact that, as frequency $\omega_1^{(\rm R)}$
is increased, the local wavefunction in the right minimum penetrates deeper into the barrier.
In this region, the approximate $n$-mode representation of the potential used in
the VCI calculations begins to deviate from the actual potential, which introduces an error
in the local energies.

The TS in the avoided crossing is reproduced with great accuracy, shown as the minima in the lower
panel in Figure \ref{fig_3}, with the error of $5 \%$.
Errors in the positions of other avoided crossings (IV-VI in Figure \ref{fig_2}), namely between
the ground state of the right minimum and the $\omega_1^{(\rm L)}$- and
$\omega_2^{(\rm L)}$-excited states in the left minimum
(frames V and VI in Figure \ref{fig_2}, respectively) become larger even using VCI energies.
The local wavefunction of the right minimum has a larger energy and penetrates deeper
into the region where the $n$-mode representation of the potential becomes unreliable.
Nevertheless, the TSs in the avoided crossings are again reproduced
accurately, which indicates that the JFI method can indeed give reliable
TM elements between different vibrational states of L/R minima and, in combination with
the VCI energies, is a useful tool for computing vibrational tunneling spectra.
Similar results were observed with the frequency $\omega_2^{(\rm R)}$ varied
(shown in Figure 3 and 4 in ESI).

Figure \ref{fig_4} shows the dependence of energy levels with the variation in the depth $d$
of the right minimum. Overall, the introduction of the energy asymmetry between the wells
results in a similar energy level pattern to that observed above. A notable difference is that,
in this case, the TSs obtained using harmonic energies are much closer to the exact QM values.
This is an artefact of the construction of the PES, in which the frequencies in the left and
right minimum are the same. As a result, the shapes of the local potentials in both minima
are similar and a large part of the error introduced by the anharmonic terms cancels out.
However, in realistic applications, it is unlikely that systems with minima of different
energies have the same L/R frequencies. The error in the position of the avoided crossing IV
is also much smaller for the harmonic energies ($\approx 2 \%$), while it is further reduced
using VCI energies ($0.7 \%$), as shown in Figure \ref{fig_5}. The error in the TS
in the avoided crossing is $12 \%$, which is comparable to the error in the case of
the frequency variation, shown in Figure \ref{fig_3}.

\subsection{MALONALDEHYDE}
\begin{figure}
\begin{center}
{\includegraphics[width=8cm]{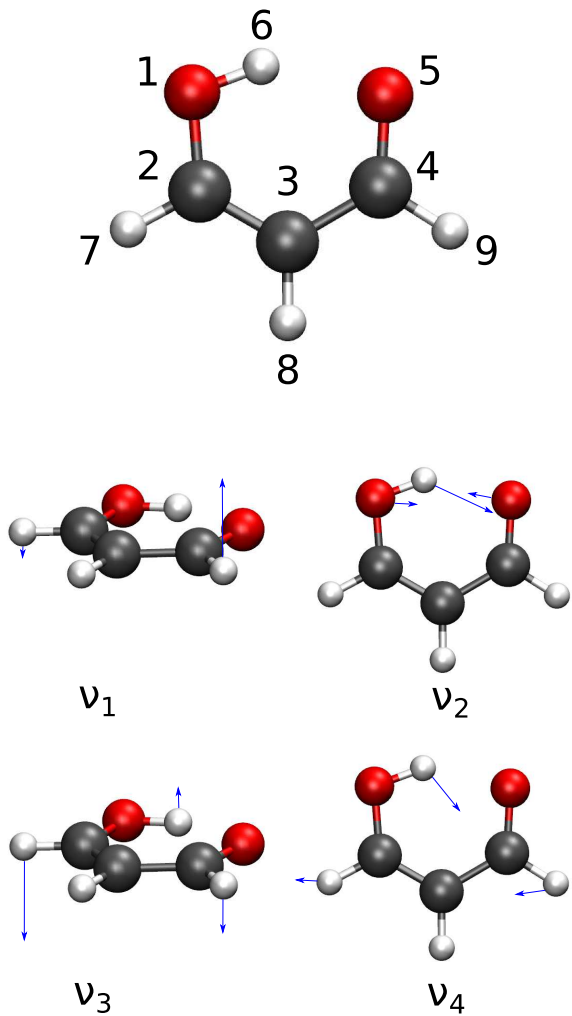}}
\caption{Annotated equilibrium geometry of malonaldehyde and schematic representation of
four lowest-frequency normal modes.}
\label{normal_modes}
\end{center}
\end{figure}

We next employ our combined approach to study the symmetric, homoisotopic malonaldehyde
on the PES developed by Wang {\em et al}\, \cite{Wang2008malonaldehydePES}. The molecule is
shown labelled in the top panel in Figure \ref{normal_modes}.
It has two equivalent wells with hydrogen 6 attached to either oxygen 1 or 5.
We study below the effect of adding additional states in the tunneling matrix.
For this purpose, vibrational energies are computed either from a $2 \times 2$ matrix involving
corresponding states in the two wells, an $8 \times 8$ matrix involving 4 local states at both
sides of the barrier, and a $16 \times 16$ matrix model. Thereby, we again calculate the local
single-well states using VSCF/VCI, while the TM matrix elements are computed using the recently
developed JFI method \cite{Erakovic2020vib}.

Malonaldehyde has been extensively studied in the past \cite{Malonald} 
and presents a benchmark system for the development of quantum dynamical
methods. Most recent calculations on the same PES using exact quantum methods
were obtained using MCTDH by Hammer and Manthe \cite{Hammer2012} and
Schr\"{o}der and Meyer \cite{Schroder2014} and show a good level of agreement
with experimental results \cite{Luttschwager2013}.
We use the results of Ref.~\onlinecite{Hammer2012} for comparison, as they 
report TSs for a number of vibrationally excited states having a
large transition dipole moment and are believed to be more accurate \cite{Schroder2014}. 

Local harmonic and VSCF/VCI energies, calculated in a $2$-mode representation
of the single-well potential, as described in Appendix C,
for the lowest 8 vibrational states, that we consider below, are shown in
Table \ref{harmonic_vci_malon}. The ground state is labelled GS, while
the excited states are labelled by the frequencies $\nu_i$ of the excited normal modes,
numbered in the order of increasing frequency in the subscript, and separated by a `$+$' sign
for multiple excitations. A noticable shift can be observed between all harmonic and VCI
energies in Table \ref{harmonic_vci_malon} due to anharmonicity, but the order in energies
remains unchanged. 
The lowest four normal modes that can get excited in the lowest 8 local
vibrational states, and that play a role in our calculations below, are depicted in
Figure \ref{normal_modes}. Higher vibrational states become more densely spaced in
energy and start to mix vibrational modes at minima. Our approach relies on being able to
uniquely define the excited normal modes at minima for each local vibrational state
considered, because the instanton wavefunctions, that are used to calculate the TM
elements that connect these states, tend to harmonic oscillator eigenstates at minima.
Moreover, a higher density of states at higher energies would require inclusion of
many additional states in the tunneling matrix, which are not known as precisely as
for the low-lying states and would thus degrade the accuracy. We limit ourselves, therefore, to
the lowest 8 local states in the studies of tunneling spectra of malonaldehyde below.

\begin{table}[hbtp]
\begin{center}
\begin{tabular}{c|c|c}
     State & Harmonic & VCI \\
     \cline{1-3}
     \multirow{2}{*}{GS} & $14950.11$ & $14682.46$\\
                        & $(0.00)$ & $(0.00)$\\
     \cline{1-3}
     \multirow{2}{*}{$\nu_1$} & $15218.68$ & $15012.65$\\
                        & $(268.57)$ & $(330.19)$\\
     \cline{1-3}
     \multirow{2}{*}{$\nu_2$} & $15245.53$ & $15042.51$\\
                        & $(295.42)$ & $(360.05)$\\
     \cline{1-3}
     \multirow{2}{*}{$\nu_3$} & $15333.29$ & $15133.95$\\
                        & $(383.17)$ & $(451.49)$\\
     \cline{1-3}
     \multirow{2}{*}{$\nu_1+\nu_1$} & $15487.25$ & $15262.05$\\
                        & $(537.14)$ & $(579.59)$\\
     \cline{1-3}
     \multirow{2}{*}{$\nu_4$} & $15472.20$ & $15281.89$\\
                        & $(522.08)$ & $(599.43)$\\
     \cline{1-3}
     \multirow{2}{*}{$\nu_2+\nu_2$} & $15540.95$ & $15318.85$\\
                        & $(590.83)$ & $(636.40)$\\
     \cline{1-3}
     \multirow{2}{*}{$\nu_1+\nu_2$} & $15514.10$ & $15336.16$\\
                        & $(563.99)$ & $(653.70)$ \\
     \cline{1-3}
\end{tabular}
\caption{Harmonic and VCI energies in $\rm{cm}^{-1}$ of the first 8 local
vibrational states of malonaldehyde labelled by the excited normal mode
frequencies. Energies relative to the local ground state are given in parentheses.}
\label{harmonic_vci_malon}
\end{center}
\end{table}

The TM elements in the $\mathbf{h}$ matrix that connect the two sets of local states
in the L and R wells are calculated using the JFI method and listed in Table \ref{tm_malon}.
Both minima of malonaldehyde belong to the $C_s$ symmetry group and its local vibrational states
can be classified according to the irreducible representation of the excited normal mode
$\nu_i$ at the minimum. The $C_s$ symmetry is preserved along the MAP, so that the TM elements
that connect normal modes of different symmetry vanish exactly, as seen in Table \ref{tm_malon}.
 
In a $2 \times 2$ matrix model, only the diagonal elements of $\mathbf{h}$ matrix are used and
the degenerate vibrational states of L/R wells are split into doublets. Equivalent results
are obtained using the first-order perturbation theory for degenerate states, yielding
the TS of $\Delta_i = 2h_{ii}$.
Energies of the GS and the first 3 excited states obtained in this manner already show a good
agreement with the MCTDH results of Ref.~\onlinecite{Hammer2012},
as can be seen in Table \ref{energies_malon} (from the second and the last column).
The vibrational states are numbered in order of increasing energy in Table \ref{energies_malon}.
The wavefunction content, obtained from the eigenvectors of the TM, is listed in
Table \ref{eigenvec_malon} and can be used to identify states in Table \ref{energies_malon}
in terms of the excited normal modes.

The TSs for the GS and the singly excited modes $\nu_{1-4}$ in the $2 \times 2$ TM model
are obtained as $\Delta(\rm GS) = 24.60$ $\rm{cm}^{-1}$, $\Delta(\nu_1) = 13.40$ $\rm{cm}^{-1}$,
$\Delta(\nu_2)=88.40$ $\rm{cm}^{-1}$, $\Delta(\nu_3) = 17.06$ $\rm{cm}^{-1}$, 
$\Delta(\nu_4) = 15.64$ $\rm{cm}^{-1}$. The MCTDH results \cite{Hammer2012} for the TSs in
the same states are $\Delta(\rm GS) = 23.5$ $\rm{cm}^{-1}$, $\Delta(\nu_1) = 6.7$ $\rm{cm}^{-1}$,
$\Delta(\nu_2)= 69.9$ $\rm{cm}^{-1}$, $\Delta(\nu_3) = 16.3$ $\rm{cm}^{-1}$, 
$\Delta(\nu_4) = 18.8$ $\rm{cm}^{-1}$. Differences in TSs, apart from the $\nu_1$- and
$\nu_2$-excited modes are well within the estimated error of the MCTDH calculations, which
validates the accuracy of our approach. The $\nu_2$ mode corresponds to the longitudinal mode
as it lies parallel to the MAP at minima. The excitation of this mode effectively lowers the
barrier of the tunneling motion and the instanton theory is known to overestimate TSs in
the shallow tunneling regime \cite{tunnel,Erakovic2020vib}. The wavefunction also penetrates
deeper into the barrier where the anharmonic effects are larger and the VCI energies degrade
as a result. Thus the accuracy in absolute energies in Table \ref{energies_malon} is also expected
to be affected for these states. The large increase in the TS for the excitation of the
longitudinal mode is, however, expected \cite{Erakovic2020vib} as confirmed by our results.
The TS for the $\nu_1$ mode is overestimated by a factor of two. This is most likely due to
the anharmonicity along this normal mode, indicated by the large difference between the harmonic
(268.57 ${\rm cm}^{-1}$) and VCI (330.19 ${\rm cm}^{-1}$) energies. Since the TS for the pair of states
is significantly suppressed compared to the GS, the frequency and energy in its direction change
substantially along the MAP. Therefore, if the anharmonicity also changes significantly, it could
cause the observed discrepancy. As an aside, we also note here that the other TSs computed
using MCTDH in Ref.~\onlinecite{Hammer2012}, which do not result in the mixture of normal modes
at minima, are $\Delta(\nu_5)=21.1$ $\rm{cm}^{-1}$, $\Delta(\nu_7) = 33.3$ $\rm{cm}^{-1}$, 
$\Delta(\nu_8) = 14.6$ $\rm{cm}^{-1}$ and $\Delta(\nu_{11}) = 19.5$ $\rm{cm}^{-1}$, and are in good
agreement with the values we obtain using JFI theory as
$\Delta(\nu_5)=24.4$ $\rm{cm}^{-1}$, $\Delta(\nu_7) = 39.5$ $\rm{cm}^{-1}$, 
$\Delta(\nu_8) = 15.6$ $\rm{cm}^{-1}$ and $\Delta(\nu_{11}) = 22.1$ $\rm{cm}^{-1}$.

\begin{table*}[hbtp]
\begin{center}
\begin{tabular}{|c|c|c|c|c|c|c|c|c|}
 \hline
  & $\rm{GS}^{(\rm R)}$ & $\nu_1^{(\rm R)}$ & $\nu_2^{(\rm R)}$ & $\nu_3^{(\rm R)}$ & $(\nu_1+\nu_1)^{(\rm R)}$ & 
    $\nu_4^{(\rm R)}$ & $(\nu_2+\nu_2)^{(\rm R)}$ & $(\nu_1+\nu_2)^{(\rm R)}$ \\
 \cline{1-9}
           $\rm{GS}^{(\rm L)}$ & $   -12.30$ & $   0.00$ & $   -21.94$ & $   0.00$ & $    -4.98$ & $    -4.62$ & $   -25.53$ & $   0.00$ \\
 \cline{1-9}
        $\nu_1^{(\rm L)}$ & $   0.00$ & $    -6.70$ & $   0.00$ & $   6.85$ & $   0.00$ & $   0.00$ & $   0.00$ & $   -11.95$ \\
 \cline{1-9}
        $\nu_2^{(\rm L)}$ & $  -21.94$ & $    0.00$ & $  -44.20$ & $    0.00$ & $   -8.87$ & $   -7.54$ & $  -55.97$ & $    0.00$ \\
 \cline{1-9}
        $\nu_3^{(\rm L)}$ & $   0.00$ & $   6.86$ & $   0.00$ & $   8.53$ & $    0.00$ & $   0.00$ & $   0.00$ & $  12.22$ \\
 \cline{1-9}
$(\nu_1+\nu_1)^{(\rm L)}$ & $    -4.98$ & $   0.00$ & $    -8.88$ & $    0.00$ & $    -4.84$ & $    -1.87$ & $    -9.14$ & $   0.00$ \\
 \cline{1-9}
        $\nu_4^{(\rm L)}$ & $    -4.61$ & $    0.00$ & $    -7.53$ & $   0.00$ & $    -1.87$ & $   7.82$ & $    -8.14$ & $    0.00$ \\
 \cline{1-9}
$(\nu_2+\nu_2)^{(\rm L)}$ & $   -25.53$ & $   0.00$ & $   -55.97$ & $   0.00$ & $    -9.14$ & $    -8.15$ & $   -75.86$ & $   0.00$ \\
 \cline{1-9}
$(\nu_1+\nu_2)^{(\rm L)}$ & $    0.00$ & $  -11.95$ & $    0.00$ & $   12.22$ & $    0.00$ & $    0.00$ & $    0.00$ & $  -21.31$ \\
 \hline
\end{tabular}
\caption{Tunneling matrix elements connecting the first 8 local vibrational states of different minima in malonaldehyde.}
\label{tm_malon}
\end{center}
\end{table*}

\begin{table}[hbtp]
\begin{center}
\scalebox{0.95}{
\begin{tabular}{c|c|c|c|c}
     No. & $E^{(\rm pairs)}$ & $E^{(4)}$ & $E^{(8)}$ & $E^{(\rm{MCTDH})}$\\
     \hline
     1 & 14670.15 & 14668.69 & 14667.08 & 14671.3 \\
     2 & 14694.76 & 14693.54 & 14692.76 & 14694.8 \\
     3 & 14998.31 & 14999.77 & 14987.74 & 14941.5 \\
     4 & 15005.95 & 15005.60 & 15005.09 & 15008.2 \\
     5 & 15019.35 & 15018.91 & 15018.54 & 15014.9 \\
     6 & 15086.70 & 15087.92 & 15077.14 & 15005.4 \\
     7 & 15125.42 & 15125.86 & 15125.14 & 15108.3 \\
     8 & 15142.47 & 15142.82 & 15142.04 & 15124.6 \\
     9 & 15243.00 & - & 15249.04 & - \\
     10 & 15257.21 & - & 15263.41 & - \\
     11 & 15266.89 & - & 15266.12 & - \\
     12 & 15274.07 & - & 15273.84 & 15249.6 \\
     13 & 15289.71 & - & 15291.11 & 15268.4 \\
     14 & 15314.85 & - & 15316.14 & - \\
     15 & 15357.47 & - & 15358.55 & - \\
     16 & 15394.71 & - & 15407.27 & - \\
\end{tabular}}
\caption{Vibrational energy levels of malonaldehyde obtained using a combined VCI/instanton approach.
$E^{(\rm pairs)}$, $E^{(4)}$ and $E^{(8)}$ are energies obtained from the $2\times2$, $8\times8$ and
$16 \times 16$ matrix models, respectively, as explained in the text.
$E^{(\rm{MCTDH})}$ are MCTDH energies from Ref.~\onlinecite{Hammer2012}.}
\label{energies_malon}
\end{center}
\end{table}

We next consider construcing the tunneling matrix using 4 local states in each well. This takes into
account interactions between the doublets considered above, whereby only the states of the same symmetry
interact. If the states of the same symmetry are well separated with respect
to the size of their TM element, the shift in energy can also be computed using the second-order
perturbation theory.
When 4 local states are taken into account in the $8 \times 8$ TM model, slight shifts are observed
in the GS and $\nu_2$-doublets in Figure \ref{fig_6} (left-side spectrum).
The absolute energies change by $1.22-1.46$ $\rm{cm}^{-1}$, while perturbation theory gives
the shift of $1.34$ $\rm{cm}^{-1}$. However, the change in the TS is negligible.

In the $16 \times 16$ TM model, consisting of 8 local states in each well,
a strong interaction with the doubly-excited $(\nu_2 + \nu_2)$ mode causes
a significant shift in the energies of the GS and the $\nu_2$-excited doublets
as well as their splittings. The TSs change from $24.85$ $\rm{cm}^{-1}$
to $25.68$ $\rm{cm}^{-1}$ and from $88.15$ $\rm{cm}^{-1}$ to $89.4$ $\rm{cm}^{-1}$, which
can clearly be observed in Figure \ref{fig_6} (right-side spectrum).
A particularly strong mixing also occurs between the doubly-excited
$(\nu_2+\nu_2)$ mode and the doubly-excited $(\nu_1 + \nu_1)$ mode, for which
the lower levels in the doublets are very close in energy
($14.21$ $\rm{cm}^{-1}$) and they interact strongly ($h=9.14$ $\rm{cm}^{-1}$
in Table \ref{tm_malon}). The mixing results in visible changes in the dominant
coefficients of TM eigenvectors in Table \ref{eigenvec_malon} and leads to
observable energy shifts.
Furthermore, singly-excited $\nu_4$ mode interacts and mixes with
the doubly-excited $(\nu_1+\nu_1)$ mode, which results in the change of
its TS from $15.64$ $\rm{cm}^{-1}$ to $17.27$ $\rm{cm}^{-1}$, which is in closer
agreement with the MCTDH value of $18.8$ $\rm{cm}^{-1}$.
Finally, we remark that the TS of the doubly-excited $(\nu_1+\nu_2)$ state
amounts to $42.62$ ${\rm cm}^{-1}$ which is in good agreement with
$49.5$ ${\rm cm}^{-1}$ obtained by Schr\"{o}der and Meyer \cite{Schroder2014}.

The above results clearly show that the interactions of different vibrational states can have
a non-negligible effect, both, on the absolute values of the vibrational energies and
on the values of the tunneling splittings. This effect is especially pronounced
if two or more states of the same symmetry are close in energy and if the TM elements
that connect them are large. This scenario is expected to play a significant role
in the higher vibrationally excited states, where the density of states becomes larger
and the interactions increase due to the presence of multiple excitations.

\begin{table}[hbtp]
\begin{center}
\scalebox{0.9}{
\begin{tabular}{c|c|c}
     No. & Pairs & (8) \\
     \cline{1-3}
     \multirow{2}{*}{1} & $0.707\ket{\rm{GS}^{(\rm{L})}}$ & $0.704\ket{\rm{GS}^{(\rm{L})}}$\\
                        & $0.707\ket{\rm{GS}^{(\rm{R})}}$ & $0.704\ket{\rm{GS}^{(\rm{R})}}$\\
     \cline{1-3}
     \multirow{2}{*}{2} & $0.707\ket{\rm{GS}^{(\rm{L})}}$ & $0.704\ket{\rm{GS}^{(\rm{L})}}$\\
                        & $-0.707\ket{\rm{GS}^{(\rm{R})}}$ & $-0.704\ket{\rm{GS}^{(\rm{R})}}$\\
     \cline{1-3}
     \multirow{2}{*}{3} & $0.707\ket{\nu_2^{(\rm{L})}}$ & $0.688\ket{\nu_2^{(\rm{L})}}$\\
                        & $0.707\ket{\nu_2^{(\rm{R})}}$ & $0.688\ket{\nu_2^{(\rm{R})}}$\\
     \cline{1-3}
     \multirow{2}{*}{4} & $0.707\ket{\nu_1^{(\rm{L})}}$ & $0.706\ket{\nu_1^{(\rm{L})}}$\\
                        & $0.707\ket{\nu_1^{(\rm{R})}}$ & $0.706\ket{\nu_1^{(\rm{R})}}$\\
     \cline{1-3}
     \multirow{2}{*}{5} & $0.707\ket{\nu_1^{(\rm{L})}}$ & $0.705\ket{\nu_1^{(\rm{L})}}$\\
                        & $-0.707\ket{\nu_1^{(\rm{R})}}$ & $-0.705\ket{\nu_1^{(\rm{R})}}$\\
     \cline{1-3}
     \multirow{2}{*}{6} & $0.707\ket{\nu_2^{(\rm{L})}}$ & $0.695\ket{\nu_2^{(\rm{L})}}$\\
                        & $-0.707\ket{\nu_2^{(\rm{R})}}$ & $-0.695\ket{\nu_2^{(\rm{R})}}$\\
     \cline{1-3}
     \multirow{2}{*}{7} & $0.707\ket{\nu_3^{(\rm{L})}}$ & $0.705\ket{\nu_3^{(\rm{L})}}$\\
                        & $-0.707\ket{\nu_3^{(\rm{R})}}$ & $-0.705\ket{\nu_3^{(\rm{R})}}$\\
     \cline{1-3}
     \multirow{2}{*}{8} & $0.707\ket{\nu_3^{(\rm{L})}}$ & $0.704\ket{\nu_3^{(\rm{L})}}$\\
                        & $0.707\ket{\nu_3^{(\rm{R})}}$ & $0.704\ket{\nu_3^{(\rm{R})}}$\\
     \cline{1-3}
     \multirow{4}{*}{9} & & $0.447\ket{(\nu_1+\nu_1)^{(\rm{L})}}$\\
                        & $0.707\ket{(\nu_2+\nu_2)^{(\rm{L})}}$ & $0.522\ket{(\nu_2+\nu_2)^{(\rm{L})}}$\\
                        & $0.707\ket{(\nu_2+\nu_2)^{(\rm{R})}}$ & $0.447\ket{(\nu_1+\nu_1)^{(\rm{R})}}$\\
                        & & $0.522\ket{(\nu_2+\nu_2)^{(\rm{R})}}$\\
     \cline{1-3}
     \multirow{4}{*}{10} & & $0.547\ket{(\nu_1+\nu_1)^{(\rm{L})}}$\\
                        & $0.707\ket{(\nu_1+\nu_1)^{(\rm{L})}}$ & $-0.436\ket{(\nu_2+\nu_2)^{(\rm{L})}}$\\
                        & $0.707\ket{(\nu_1+\nu_1)^{(\rm{R})}}$ & $0.547\ket{(\nu_1+\nu_1)^{(\rm{R})}}$\\
                        & & $-0.436\ket{(\nu_2+\nu_2)^{(\rm{R})}}$\\
     \cline{1-3}
     \multirow{2}{*}{11} & $0.707\ket{(\nu_1+\nu_1)^{(\rm{L})}}$ & $0.693\ket{(\nu_1+\nu_1)^{(\rm{L})}}$\\
                        & $-0.707\ket{(\nu_1+\nu_1)^{(\rm{R})}}$ & $-0.693\ket{(\nu_1+\nu_1)^{(\rm{R})}}$\\
     \cline{1-3}
     \multirow{2}{*}{12} & $0.707\ket{\nu_4^{(\rm{L})}}$ & $0.693\ket{\nu_4^{(\rm{L})}}$\\
                        & $-0.707\ket{\nu_4^{(\rm{R})}}$ & $-0.693\ket{\nu_4^{(\rm{R})}}$\\
     \cline{1-3}
     \multirow{2}{*}{13} & $0.707\ket{\nu_4^{(\rm{L})}}$ & $0.693\ket{\nu_4^{(\rm{L})}}$\\
                        & $0.707\ket{\nu_4^{(\rm{R})}}$ & $0.693\ket{\nu_4^{(\rm{R})}}$\\
     \cline{1-3}
     \multirow{2}{*}{14} & $0.707\ket{(\nu_1+\nu_2)^{(\rm{L})}}$ & $0.705\ket{(\nu_1+\nu_2)^{(\rm{L})}}$\\
                        & $0.707\ket{(\nu_1+\nu_2)^{(\rm{R})}}$ & $0.705\ket{(\nu_1+\nu_2)^{(\rm{R})}}$\\
     \cline{1-3}
     \multirow{2}{*}{15} & $0.707\ket{(\nu_1+\nu_2)^{(\rm{L})}}$ & $0.706\ket{(\nu_1+\nu_2)^{(\rm{L})}}$\\
                        & $-0.707\ket{(\nu_1+\nu_2)^{(\rm{R})}}$ & $-0.706\ket{(\nu_1+\nu_2)^{(\rm{R})}}$\\
     \cline{1-3}
     \multirow{2}{*}{16} & $0.707\ket{(\nu_2+\nu_2)^{(\rm{L})}}$ & $0.691\ket{(\nu_2+\nu_2)^{(\rm{L})}}$\\
                        & $-0.707\ket{(\nu_2+\nu_2)^{(\rm{R})}}$ & $-0.691\ket{(\nu_2+\nu_2)^{(\rm{R})}}$\\
     \cline{1-3}

\end{tabular}}
\caption{Dominant configurations of vibrational states of malonaldehyde, obtained as 
the eigenvectors of tunneling matrix in the $2 \times 2$ (pairs) and $16 \times 16$ (8-state)
models, as described in the text.}
\label{eigenvec_malon}
\end{center}
\end{table}

\subsection{PARTIALLY DEUTERATED MALONALDEHYDE}
In the previous Subsection, we have learned what accuracy one might expect in
the calculation of the tunneling spectra of malonaldehyde through comparison with
the exact QM results. We now consider the partially deuterated (PD) malonaldehyde,
where hydrogen in the position 7/9 is substituted by deuterium (see Figure \ref{normal_modes})
and the system in no longer symmetric. Since deuterium is not placed in equivalent positions
in the two minima, their local vibrational frequencies and energies are no longer equal,
even though the PES remains unchanged.
The particular choice of deuteration was chosen for our study because the mixing angle
in its GS was determined experimentally by Baughcum {\em et al}\, \cite{Baughcum1981malonaldehyde}
and the TS by Jahr {\em et al}\, \cite{Jahr2020} using RPI method.
Furthermore, the size of the relative energy shifts between the left and right minimum is
comparable to the size of the TM elements, which makes the system interesting in that
both the VCI energies and the instanton TM elements are expected to make a significant
contribution to the TSs in this system.

\begin{table}[hbtp]
\begin{centering}
\small
\scalebox{0.9}{
\begin{tabular}{c|c|c|c|c}
    \multirow{2}{*}{State}  & \multicolumn{2}{c|}{Harmonic} & \multicolumn{2}{c}{VCI} \\
    \cline{2-5}
      & D7 & D9 & D7 & D9\\
     \cline{1-5}
     \multirow{2}{*}{GS} & $14228.18$ & $14253.67$ & $13978.19$ & $14013.04$\\
                        & $(0.00)$ & $(25.49)$ & $(0.00)$ & $(34.85)$\\
     \cline{1-5}
     \multirow{2}{*}{$\nu_1$} & $14492.55$ & $14492.10$ & $14298.70$ & $14311.75$\\
                        & $(264.37)$ & $(263.92)$ & $(320.51)$ & $(333.56)$\\
     \cline{1-5}
     \multirow{2}{*}{$\nu_2$} & $14522.65$ & $14547.57$ & $14327.02$ & $14361.38$\\
                        & $(294.47)$ & $(319.39)$ & $(348.83)$ & $(383.19)$\\
     \cline{1-5}
     \multirow{2}{*}{$\nu_3$} & $14568.85$ & $14626.56$ & $14384.49$ & $14444.96$\\
                        & $(340.67)$ & $(398.38)$ & $(406.30)$ & $(466.76)$\\
     \cline{1-5}
     \multirow{2}{*}{$\nu_1+\nu_1$} & $14756.92$ & $14546.67$ & $14950.11$ & $14543.15$\\
                        & $(528.74)$ & $(502.34)$ & $(568.48)$ & $(564.96)$\\
     \cline{1-5}
     \multirow{2}{*}{$\nu_4$} & $14744.31$ & $14769.16$ & $14562.35$ & $14595.52$\\
                        & $(516.13)$ & $(540.98)$ & $(584.16)$ & $(617.32)$\\
     \cline{1-5}
     \multirow{2}{*}{$\nu_2+\nu_2$} & $14817.13$ & $14841.47$ & $14601.27$ & $14637.01$\\
                        & $(588.95)$ & $(613.29)$ & $(623.08)$ & $(658.82)$\\
     \cline{1-5}
     \multirow{2}{*}{$\nu_1+\nu_2$} & $14787.02$ & $14786.00$ & $14614.31$ & $14624.21$\\
                        & $(558.84)$ & $(557.82)$ & $(636.12)$ & $(646.02)$\\
     \cline{1-5}
\end{tabular}}
\caption{Harmonic and VCI energies in $\rm{cm}^{-1}$ of the first 8 local
vibrational states of partially deuterated malonaldehyde labelled by
the excited normal mode frequencies. Energies relative to the local ground state
of the D7 minimum are given in parentheses.}
\label{harmonic_vci_pdm}
\end{centering}
\end{table}

\begin{table*}[hbtp]
\begin{center}
\scalebox{0.95}{
\begin{tabular}{|c|c|c|c|c|c|c|c|c|}
 \hline
  & $\rm{GS}^{(\rm D9)}$ & $\nu_1^{(\rm D9)}$ & $\nu_2^{(\rm D9)}$ & $\nu_3^{(\rm D9)}$ & $(\nu_1+\nu_1)^{(\rm D9)}$ & $\nu_4^{(\rm D9)}$ & $(\nu_2+\nu_2)^{(\rm D9)}$ & $(\nu_1+\nu_2)^{(\rm D9)}$ \\
 \cline{1-9}
        \multirow{2}{*}{$\rm{GS}^{(\rm D7)}$} & $   -12.32$ & $    0.00$ & $  -21.95$ & $    0.00$ & $    -5.04$ & $    -4.63$ & $   -25.54$ & $   0.00$ \\
                                              & $   (0.005)$ & $   (0.00)$ & $(0.133)$ & $(0.00)$ & $(0.042)$ & $(0.034)$ & $(0.343)$ & $(0.00)$ \\
 \cline{1-9}
        \multirow{2}{*}{$\nu_1^{(\rm D7)}$} & $   0.00$ & $   -6.86$ & $    0.00$ & $    6.52$ & $   0.00$ & $   0.00$ & $   0.00$ & $   -12.22$ \\
                                              & $   (0.00)$ & $   (0.001)$ & $(0.00)$ & $(0.014)$ & $(0.00)$ & $(0.00)$ & $(0.00)$ & $(0.077)$ \\
 \cline{1-9}
        \multirow{2}{*}{$\nu_2^{(\rm D7)}$} & $  -21.89$ & $   0.00$ & $   -44.12$ & $   0.00$ & $   -8.95$ & $   -7.52$ & $  -55.89$ & $    0.00$ \\
                                              & $   (0.109)$ & $   (0.00)$ & $(0.031)$ & $(0.00)$ & $(0.037)$ & $(0.029)$ & $(0.452)$ & $(0.00)$ \\
 \cline{1-9}
        \multirow{2}{*}{$\nu_3^{(\rm D7)}$} & $   0.00$ & $    7.37$ & $    0.00$ & $    8.64$ & $   0.00$ & $   0.00$ & $   0.00$ & $  13.14$ \\
                                              & $   (0.00)$ & $   (0.007)$ & $(0.00)$ & $(0.007)$ & $(0.00)$ & $(0.00)$ & $(0.00)$ & $(0.054)$ \\
 \cline{1-9}
        \multirow{2}{*}{$(\nu_1+\nu_1)^{(\rm D7)}$} & $    -5.16$ & $    0.00$ & $   -9.19$ & $   0.00$ & $    -4.58$ & $    -1.94$ & $    -9.45$ & $   0.00$ \\
                                                    & $   (0.042)$ & $   (0.00)$ & $(0.034)$ & $(0.00)$ & $(0.000)$ & $(0.001)$ & $(0.024)$ & $(0.00)$ \\
 \cline{1-9}
        \multirow{2}{*}{$\nu_4^{(\rm D7)}$} & $   -4.44$ & $    0.00$ & $    -7.25$ & $   0.00$ & $   -1.82$ & $    7.97$ & $   -7.84$ & $   0.00$ \\
                                            & $   (0.029)$ & $   (0.00)$ & $(0.021)$ & $(0.00)$ & $(0.001)$ & $(0.004)$ & $(0.010)$ & $(0.00)$ \\
 \cline{1-9}
        \multirow{2}{*}{$(\nu_2+\nu_2)^{(\rm D7)}$} & $   -25.42$ & $   0.00$ & $  -55.79$ & $    0.00$ & $    -9.18$ & $    -8.10$ & $   -75.67$ & $    0.00$ \\
                                                    & $   (0.304)$ & $   (0.00)$ & $(0.349)$ & $(0.00)$ & $(0.015)$ & $(0.001)$ & $(0.090)$ & $(0.00)$ \\
 \cline{1-9}
        \multirow{2}{*}{$(\nu_1+\nu_2)^{(\rm D7)}$} & $    0.00$ & $   -12.19$ & $   0.00$ & $  11.59$ & $    0.00$ & $    0.00$ & $    0.00$ & $  -21.72$ \\
                                                    & $   (0.00)$ & $   (0.072)$ & $(0.133)$ & $(0.037)$ & $(0.00)$ & $(0.00)$ & $(0.00)$ & $(0.006)$ \\
 \hline
\end{tabular}}

\caption{Tunneling matrix elements connecting the first 8 local vibrational states of different minima in 
partially deuterated malonaldehyde. Values in parentheses refer to the estimated error introduced by
the neglect of overlap between L/R local states, as explained in Appendix B.}
\label{tm_pdm}
\end{center}
\end{table*}

\begin{table}[hbtp]
\begin{center}
\begin{tabular}{c|c|c|c}
     No. & $E^{(\rm pairs)}$ & $E^{(4)}$ & $E^{(8)}$\\
     \hline
     1 & 13974.27 & 13972.91 & 13971.48 \\
     2 & 14016.96 & 14015.52 & 14014.49 \\
     3 & 14296.86 & 14298.42 & 14286.79 \\
     4 & 14295.75 & 14295.44 & 14294.91 \\
     5 & 14314.69 & 14313.95 & 14313.54 \\
     6 & 14391.55 & 14392.78 & 14381.17 \\
     7 & 14383.28 & 14384.05 & 14383.32 \\
     8 & 14446.16 & 14446.45 & 14445.64 \\
     9 & 14540.00 & - & 14537.31 \\
     10 & 14549.82 & - & 14549.32 \\
     11 & 14541.38 & - & 14556.57 \\
     12 & 14560.53 & - & 14561.08 \\
     13 & 14597.33 & - & 14598.21 \\
     14 & 14596.99 & - & 14598.33 \\
     15 & 14641.54 & - & 14642.66 \\
     16 & 14696.90 & - & 14709.18 \\
\end{tabular}
\caption{Vibrational energy levels of partially deuterated malonaldehyde obtained using a combined
VCI/instanton approach.
$E^{(\rm pairs)}$, $E^{(4)}$ and $E^{(8)}$ are energies obtained from the $2\times2$, $8\times8$ and
$16 \times 16$ matrix models, respectively, as explained in the text.}
\label{energies_pdm}
\end{center}
\end{table}

In the PD malonaldehyde, the isotopic substitution causes a significant lowering of
the zero-point energy, given in Table \ref{harmonic_vci_pdm}, from $14682.45$ $\rm{cm}^{-1}$
to $13978.19$ $\rm{cm}^{-1}$ for D7 minimum and to $14013.04$ $\rm{cm}^{-1}$ for D9 minimum.
Additionally, the excitation energies for the first 7 excited states decrease as well,
by up to $40$ $\rm{cm}^{-1}$. As a result, the vibrational states are more closely spaced,
see Figure \ref{fig_7}, and larger interstate L/R mixings are expected.

The normal modes in PD malonaldehyde are qualitatively similar to the homoisotopic
malonaldehyde, depicted in Figure \ref{normal_modes}. The ordering of local
single-well states, labelled by the excited normal mode at minimum, is also preserved
upon deuteration, with the exception of the $\ket{(\nu_1+\nu_2)^{(\rm D9)}}$ and
$\ket{(\nu_2+\nu_2)^{(\rm D9)}}$ states, which exchange order.
The TM elements, shown in Table \ref{tm_pdm}, are remarkably
similar to the homoisotopic malonaldehyde, which indicates that the wavefunctions
in the barrier region are not significantly affected by the asymmetry. The error
estimates due to the variation of the position of the dividing plane are shown in
parentheses in Table \ref{tm_pdm}, and are discussed in more detail in Appendix B.

We again consider pairwise interactions of the corresponding states in a $2 \times 2$
TM model. This is possible since the normal modes at both minima can approximately
be mapped to one another using a symmetry operation. The pairs of states are no longer
degenerate in this case, and the first-order perturbation theory cannot be used
to estimate the TSs. Instead, the TS, obtained from the eigenvalues of the TM,
is seen to be equal to the local energy difference corrected by the second-order
perturbative terms,
\begin{align}
    \nonumber
    \Delta_i &= \sqrt{(E_i^{(\rm D7)}-E_i^{(\rm D9)})^2+4h_{ii}^2} \\
             &\approx |E_i^{(\rm D7)}-E_i^{(\rm D9)}| + \frac{2h_{ii}^2}{|E_i^{(\rm D7)}-E_i^{(\rm D9)}|},
\label{2nd_corr}
\end{align}
where, in the last line of \eqn{2nd_corr}, we assumed that the TM element
$|h_{ii}| \ll |E_i^{(\rm D7)}-E_i^{(\rm D9)}|$. This assumption is certainly
violated if there are other local states which are energetically close and
coupled by the TM elements that are comparable in size.

The TM element for the GS is $12.32$ $\rm{cm}^{-1}$ using JFI method, which is
in excellent agreement with $12.4$ $\rm{cm}^{-1}$ obtained by
Jahr {\em et al}\, \cite{Jahr2020} using RPI. The mixing angle for the GS 
was estimated experimentally by Baughcum {\em et al}\, \cite{Baughcum1981malonaldehyde}
at $\phi = 41^{\circ}$. Ref.~\onlinecite{Jahr2020} estimates the angle at
$\phi = 44^{\circ}$, using local harmonic energies. Using VCI energies,
we estimate the mixing angle to be $\phi=35.3^{\circ}$, which indicates that
the anharmonicity is indeed responsible for a decrease in its value, as 
speculated by Jahr {\em et al}\, \cite{Jahr2020}.
We were also able to estimate the effect of the inclusion of other local vibrational
states on the mixing angle from the components of the TM eigenvectors
in Table \ref{eigenvec_pdm} as
\begin{equation}
\tan \phi/2 = \frac{c(\rm{GS}^{(\rm D9)})}{c(\rm{GS}^{(\rm D7)})}
\end{equation}
which gives $\phi = 36.8^{\circ}$. It thus appears that the inclusion of
additional interactions corrects the mixing angle towards
the experimental value.

Changes in the vibrational levels of the excited states in the $8 \times 8$
and the $16 \times 16$ matrix models, listed in Table \ref{energies_pdm},
are qualitatively similar to the homoisotopic malonaldehyde due
to the similarity in their TM elements. The vibrational tunneling spectrum is
shown graphically in Figure \ref{fig_7}.
One significant difference here is that some doublet states change order
of their components after the inclusion of additional vibrational states
in the model due to their proximity in energy after deuteration, as seen
in Figure \ref{fig_7}.
Another difference is the absence of symmetry in the wavefunctions
with respect to the symmetry operation that connects the minima
in the homoisotopic case. As a result, the extensions of the $2 \times 2$
model to higher dimensionality matrix models will mix both,
the lower and the higher components of doublets, with all other doublet states.
Finally, due to the proximity of vibrational states, the lower components
of the $(\nu_1+\nu_1)$, $\nu_4$ and $(\nu_2+\nu_2)$ doublets are significantly
mixed, as can be seen in Table \ref{eigenvec_pdm}.
This mixing between the states changes their energies, but it is also expected
to affect the intensity of the transition to the $11^{\rm th}$ state,
as its $\nu_4$ component (see Table \ref{eigenvec_pdm}) has a higher transition
dipole moment, being the singly excited state.

\begin{table}[hbtp]
\begin{center}
\scalebox{0.9}{
\begin{tabular}{c|c|c}
     No. & Pairs & (8) \\
     \cline{1-3}
     \multirow{2}{*}{1} & $0.953\ket{\rm{GS}^{(\rm{D7})}}$ & $0.946\ket{\rm{GS}^{(\rm{D7})}}$\\
                        & $0.303\ket{\rm{GS}^{(\rm{D9})}}$ & $0.313\ket{\rm{GS}^{(\rm{D9})}}$\\
     \cline{1-3}
     \multirow{2}{*}{2} & $-0.303\ket{\rm{GS}^{(\rm{D7})}}$ & $-0.316\ket{\rm{GS}^{(\rm{D7})}}$\\
                        & $0.953\ket{\rm{GS}^{(\rm{D9})}}$ & $0.946\ket{\rm{GS}^{(\rm{D9})}}$\\
     \cline{1-3}
     \multirow{2}{*}{3} & $0.826\ket{\nu_2^{(\rm{D7})}}$ & $0.797\ket{\nu_2^{(\rm{D7})}}$\\
                        & $0.564\ket{\nu_2^{(\rm{D9})}}$ & $0.560\ket{\nu_2^{(\rm{D9})}}$\\
     \cline{1-3}
     \multirow{2}{*}{4} & $0.919\ket{\nu_1^{(\rm{D7})}}$ & $0.914\ket{\nu_1^{(\rm{D7})}}$\\
                        & $0.394\ket{\nu_1^{(\rm{D9})}}$ & $0.401\ket{\nu_1^{(\rm{D9})}}$\\
     \cline{1-3}
     \multirow{2}{*}{5} & $-0.394\ket{\nu_1^{(\rm{D7})}}$ & $-0.403\ket{\nu_1^{(\rm{D7})}}$\\
                        & $0.919\ket{\nu_1^{(\rm{D9})}}$ & $0.909\ket{\nu_1^{(\rm{D9})}}$\\
     \cline{1-3}
     \multirow{2}{*}{6} & $-0.564\ket{\nu_2^{(\rm{D7})}}$ & $-0.573\ket{\nu_2^{(\rm{D7})}}$\\
                        & $0.826\ket{\nu_2^{(\rm{D9})}}$ & $0.794\ket{\nu_2^{(\rm{D9})}}$\\
     \cline{1-3}
     \multirow{2}{*}{7} & $0.990\ket{\nu_3^{(\rm{D7})}}$ & $0.984\ket{\nu_3^{(\rm{D7})}}$\\
                        & $-0.139\ket{\nu_3^{(\rm{D9})}}$ & $-0.138\ket{\nu_3^{(\rm{D9})}}$\\
     \cline{1-3}
     \multirow{2}{*}{8} & $0.139\ket{\nu_3^{(\rm{D7})}}$ & $0.138\ket{\nu_3^{(\rm{D7})}}$\\
                        & $0.990\ket{\nu_3^{(\rm{D9})}}$ & $0.987\ket{\nu_3^{(\rm{D9})}}$\\
     \cline{1-3}
     \multirow{4}{*}{9} & & $0.505\ket{(\nu_1+\nu_1)^{(\rm{D7})}}$\\
                        & $0.566\ket{(\nu_1+\nu_1)^{(\rm{D7})}}$ & $0.293\ket{(\nu_2+\nu_2)^{(\rm{D7})}}$\\
                        & $0.824\ket{(\nu_1+\nu_1)^{(\rm{D9})}}$ & $0.761\ket{(\nu_1+\nu_1)^{(\rm{D9})}}$\\
                        & & $-0.233\ket{(\nu_2+\nu_2)^{(\rm{D9})}}$\\
     \cline{1-3}
     \multirow{2}{*}{10} & $0.824\ket{(\nu_1+\nu_1)^{(\rm{D7})}}$ & $0.830\ket{(\nu_1+\nu_1)^{(\rm{D7})}}$\\
                        & $-0.566\ket{(\nu_1+\nu_1)^{(\rm{D9})}}$ & $-0.543\ket{(\nu_1+\nu_1)^{(\rm{D9})}}$\\
     \cline{1-3}
     \multirow{4}{*}{11} & & $0.354\ket{\nu_4^{(\rm{D7})}}$\\
                        & $0.784\ket{(\nu_2+\nu_2)^{(\rm{D7})}}$ & $0.626\ket{(\nu_2+\nu_2)^{(\rm{D7})}}$\\
                        & $0.621\ket{(\nu_2+\nu_2)^{(\rm{D9})}}$ & $-0.349\ket{(\nu_1+\nu_1)^{(\rm{D9})}}$\\
                        & & $0.537\ket{(\nu_2+\nu_2)^{(\rm{D9})}}$\\
     \cline{1-3}
     \multirow{4}{*}{12} & & $0.895\ket{\nu_4^{(\rm{D7})}}$\\
                        & $0.975\ket{\nu_4^{(\rm{D7})}}$ & $-0.299\ket{(\nu_2+\nu_2)^{(\rm{D7})}}$\\
                        & $-0.222\ket{\nu_4^{(\rm{D9})}}$ & $-0.262\ket{\nu_4^{(\rm{D9})}}$\\
                        & & $-0.161\ket{(\nu_2+\nu_2)^{(\rm{D9})}}$\\
     \cline{1-3}
     \multirow{2}{*}{13} & $0.222\ket{\nu_4^{(\rm{D7})}}$ & $0.236\ket{\nu_4^{(\rm{D7})}}$\\
                        & $0.975\ket{\nu_4^{(\rm{D9})}}$ & $0.961\ket{\nu_4^{(\rm{D9})}}$\\
     \cline{1-3}
     \multirow{2}{*}{14} & $0.782\ket{(\nu_1+\nu_2)^{(\rm{D7})}}$ & $0.778\ket{(\nu_1+\nu_2)^{(\rm{D7})}}$\\
                        & $0.624\ket{(\nu_1+\nu_2)^{(\rm{D9})}}$ & $0.622\ket{(\nu_1+\nu_2)^{(\rm{D9})}}$\\
     \cline{1-3}
     \multirow{2}{*}{15} & $-0.624\ket{(\nu_1+\nu_2)^{(\rm{D7})}}$ & $-0.623\ket{(\nu_1+\nu_2)^{(\rm{D7})}}$\\
                        & $0.782\ket{(\nu_1+\nu_2)^{(\rm{D9})}}$ & $0.780\ket{(\nu_1+\nu_2)^{(\rm{D9})}}$\\
     \cline{1-3}
     \multirow{2}{*}{16} & $-0.621\ket{(\nu_2+\nu_2)^{(\rm{D7})}}$ & $-0.612\ket{(\nu_2+\nu_2)^{(\rm{D7})}}$\\
                        & $0.784\ket{(\nu_2+\nu_2)^{(\rm{D9})}}$ & $0.764\ket{(\nu_2+\nu_2)^{(\rm{D9})}}$\\
     \cline{1-3}

\end{tabular}}
\caption{Dominant configurations of vibrational states of partially deuterated malonaldehyde,
obtained as  the eigenvectors of tunneling matrix in the $2 \times 2$ (pairs) and $16 \times 16$ (8-state)
models, as described in the text.}
\label{eigenvec_pdm}
\end{center}
\end{table}

\section{Conclusions}
We applied a combination of VCI and instanton theory to
calculate vibrational tunneling spectra of some exemplary
double-well systems in full dimensionality at a much reduced
computational cost in comparison with the exact QM methods.
The VCI method was used to compute the single-well
vibrational spectra, while the recently developed instanton
method was used to determine the wavefunctions inside the barrier
that separates the wells at a comparatively negligible computational
cost. The interaction between the states of different wells was
obtained from the Herring formula evaluated at a dividing surface
inside the barrier. Herring formula was rederived in
an extended $N\times N$ matrix model ($N>2$) and the size of
the associated leading error term was analysed.

The accuracy of our approach was first tested on a model 2D system.
It was shown that the JFI method can be used to compute
TM elements that connect states in inequivalent wells and that
have excitations in different normal modes.
The energy levels of an asymmetric system exhibit avoided
crossings with the variation of frequency or depth of
one well relative to the other. The VCI calculation of local
energies proved to be necessary in order to reproduce the exact QM
results with high accuracy.
The method was then tested on malonadehyde in full dimensionality,
where good agreement was achieved with the exact MCTDH results in
the absolute energies and the splittings. It was shown that
the extension of the standard $2 \times 2$ model to include more
states can influence the vibrational energies. The results are not
affected dramatically in the case of malonaldehyde, but it was shown
that the model is able to accommodate the additional vibrational states
in the systems where they lie close in energy.

Finally, the method was used to calculate the vibrational
spectrum of the low-lying states in partially deuterated malonaldehyde,
which is near the computational limit of the presently available exact
QM methods.
The ground state mixing angle was compared to the experiment and
the influence of including additional vibrational states was shown
to affect the angle and the order of some states in the spectrum.

The method is expected to perform well for mid-sized molecules,
where rotational motion, which is neglected in this work, can be separated
from the tunneling dynamics, and for moderately anharmonic systems with
high barriers and, consequently, small tunneling splittings. It is exactly
in these circumstances that the exact QM methods come at a prohibitive
computational cost. The developed combined approach can be used to calculate
the low-lying vibrational spectra in systems with arbitrary number
of wells, which are not necessarily related by symmetry. This makes
the method particularly suitable to the studies of clusters, e.g., for
the assignment of spectra in water clusters, which feature multiple
minima and high barriers in their bifurcation dynamics (where hydrogen
bonds are broken and reformed).
The computational cost of our approach is concentrated in solving
the single-well spectra separately. The instanton theory can also be combined
with other high-level methods, instead of VCI, and the combined, dual-level,
approach can be used to calculate tunneling spectra in general
multidimensional asymmetric well systems, beyond molecular applications
and chemistry.

\begin{acknowledgments}
This work was fully supported by Croatian Science Foundation Grant No.~IP-2020-02-9932.
\end{acknowledgments}

\appendix
\section{MAP invariance with respect to the addition of a switching function}
We show below that the addition of a function $f(S)$, which only depends on
the coordinate $S$ along the path, to the potential,
\begin{equation}
\tilde{V}({\mathbf x}) = V({\mathbf x})-f(S),
\end{equation}
does not change the shape of the MAP, but only scales the imaginary time
parameter $\tau$. The statement is valid if the function $f$ does not change
the position or shape of the minima, i.e., it satisfies
\begin{align}
    \nonumber
    \lim_{S \to 0 (S_{\rm tot})} f(S) = 0, \\
    \nonumber
    \lim_{S \to 0 (S_{\rm tot})} f'(S) = 0, \\
    \lim_{S \to 0 (S_{\rm tot})} f''(S) = 0,
\label{f_conditions}
\end{align}
and is negligible in comparison with the potential
in the region between the minima, $f(S) \ll V(S)$.

Path ${\mathbf x}(\tau)$ is the characteristic for potential $V$ and it satisfies
\begin{align}
    \nonumber
    \frac{d^2}{d \tau^2} {\mathbf x}(\tau) = \nabla V. \\
    \frac{d S}{d \tau} = p_0 = \sqrt{2V},
\end{align}
where $S$ is the arc length distance along the path.
We define the scaled imaginary time as
\begin{equation}
    d\tilde{\tau} = \frac{d \tau}{\sqrt{1-\frac{f(S)}{V}}}.
\end{equation}
The scaled momentum then becomes
\begin{equation}
    \tilde{p}_0 = \frac{d S}{d \tilde{\tau}} = \sqrt{2(V-f(S))} = \sqrt{2\tilde{V}}.
\end{equation}
The momentum vector transforms as
\begin{align}
    \nonumber
    \frac{d}{d \tilde{\tau}} {\mathbf x}(\tilde{\tau}) = \frac{d {\mathbf x}}{d \tau} \sqrt{1-\frac{f(S)}{V}}, \\
    \tilde{\mathbf p}_0 = {\mathbf p}_0 \sqrt{1-\frac{f(S)}{V}},
\end{align}
whereas the acceleration becomes
\begin{align}
    \nonumber
    \frac{d^2}{d \tilde{\tau}^2} {\mathbf x}(\tilde{\tau}) &= \nabla (V-f(S)) + \frac{f(S)}{V} \left( \frac{{\mathbf p}_0}{p_0} \frac{d}{dS}V - \nabla V \right), \\
    \frac{d^2}{d \tilde{\tau}^2} {\mathbf x}(\tilde{\tau}) &= \nabla \tilde{V} - \frac{f(S)}{V} \left( \nabla V \right)_{\perp},
\label{trans_char}
\end{align}
where the symbol $\perp$ in the subscript denotes the component of the vector
that is perpendicular to the path. Now, if the function $f(S)$ satisfies
the conditions in \eqn{f_conditions} and is significantly smaller than
the potential, the second term on the right hand side of \eqn{trans_char}
is small everywhere on the path in comparison to the gradient of the potential
and can be ignored. \eqn{trans_char} then takes the form of
the equation of characteristic, but on the modified potential $\tilde{V}$.

\section{Dependence of tunneling matrix elements on the position of the dividing plane}
TM elements between the states with, at most, one excitation ($\nu$, $\nu'=0-1$)
can be expressed as
\begin{align}
    \nonumber
    h_{\nu \nu'}= &
    \sqrt{\frac{\sqrt{\det' {\mathbf A}_0^{(\rm L)} \det' {\mathbf A}_0^{(\rm R)}}}{{\pi \det' \bar{{\mathbf A}}}}} 
    \frac{p_0^{(\rm L)}+p_0^{(\rm R)}}{2} \\
    &\times \left( F^{(\rm L)} \right)^{\nu} \left( F^{(\rm R)} \right)^{\nu'} 
    \nonumber
    \sqrt{\frac{\left( 2\omega_{\rm e}^{(\rm L)} \right)^{\nu} \left( 2\omega_{\rm e}^{(\rm R)} \right)^{\nu'}}
     {(2\nu-1)!!(2\nu'-1)!!}} \\
\nonumber
    &\times {\rm e}^{-\int_0^{S_{\rm cp}} p_0^{(\rm L)}d S -\int_{S_{\rm cp}}^{S_{\rm tot}} p_0^{(\rm R)}d S} \\
    &\times {\rm e}^{-\frac{1}{2}\int_0^{S_{\rm cp}} \frac{\rm{Tr} ({\mathbf A}^{(\rm L)} - {\mathbf A}_0^{(\rm L)})}{p_0^{(\rm L)}} dS 
    -\frac{1}{2}\int_{S_{\rm cp}}^{S_{\rm tot}}
     \frac{\rm{Tr} ({\mathbf A}^{(\rm R)} - {\mathbf A}_0^{(\rm R)})}{p_0^{(\rm R)}} dS},
\label{hf_ij}
\end{align}
where the contribution of $\mathbf U$ terms in \eqn{h_ij} has been neglected. It is known that,
unless these terms exclusively contribute to the TM elements, they introduce the dependence of
the TM element $h_{\nu \nu'}$ on the connection point \cite{Erakovic2020vib}.
Therefore, we omit them from the treatment here to separate the effect of $\mathbf U$ terms from the effect
of asymmetry in the dependence of the TM element with the variation of the connection point $S_{\rm cp}$.
The case of multiple excitations, as well as that of the $\mathbf U$ terms, can be derived in an analogous
fashion.

Dependence on the position of the connection point can be determined by differentiating the TM element
in \eqn{hf_ij} with respect to $S_{\rm cp}$. Small changes in the connection point correspond to
the addition of function $f(S)$ of the following form
\begin{equation}
    f(S) = \left\{ 
    \begin{array}{r l}
        -d  & S_{\rm cp}-\varepsilon < S < S_{\rm cp} \\
        +d  & S_{\rm cp} < S < S_{\rm cp}+\varepsilon \\
        0  & {\rm elsewhere}
    \end{array}
    \right.
\end{equation}
to the potential in the action integral, where $\varepsilon$ denotes the change in the connection point.
As discussed in Appendix A, the addition of $f$ does not change the shape of the path. The differentiation
can thus be performed on the same path. Furthermore, if the connection point is located deep inside
the barrier, we can approximate
\begin{equation}
    p_0^{(\rm L)} \approx p_0^{(\rm R)} = \bar{p}_0.
\end{equation}
Following the differentiation, as described in Ref.~\onlinecite{Erakovic2020vib}, we obtain
\begin{align}
    \nonumber
    \frac{d}{dS_{\rm cp}} h_{\nu \nu'} = & \frac{h_{\nu \nu'}}{\bar{p}_0} \left[ 2\frac{d \bar{p}_0}{d S_{\rm cp}} + \omega_{\rm e}^{(\rm L)} \delta_{\nu,1} - \omega_{\rm e}^{(\rm R)} \delta_{\nu',1}  + \right. \\
    \nonumber
    &\left. \bar{p}_0(p_0^{(\rm R)}-(p_0^{(\rm L)}) + \frac{1}{2} {\rm Tr} \left( {\mathbf A}_{\perp}^{(\rm L)}-{\mathbf A}_{\perp}^{(\rm R)}\right) -  \right. \\
    \nonumber
    &\left. \frac{1}{2} {\rm Tr} \left( {\mathbf A}^{(\rm L)}-{\mathbf A}^{(\rm R)}\right) + \frac{1}{2} {\rm Tr} \left( {\mathbf A}_0^{(\rm L)}-{\mathbf A}_0^{(\rm R)}\right)\right] \\
    \nonumber
    \frac{d}{dS_{\rm cp}} h_{\nu \nu'} = & \frac{h_{\nu \nu'}}{\bar{p}_0} \left[ \omega_{\rm e}^{(\rm L)} \delta_{\nu,1} - \omega_{\rm e}^{(\rm R)} \delta_{\nu',1} + \right. \\ 
    &\left.\frac{1}{2} \left(p_0^{(\rm R)\mbox{}^{\scriptstyle 2}}-p_0^{(\rm L)^{\scriptstyle 2}} \right) \right. 
    \nonumber
    \left.+\frac{1}{2} {\rm Tr} \left( {\mathbf A}_0^{(\rm L)}-{\mathbf A}_0^{(\rm R)}\right) \right] \\
    \nonumber
    \frac{d}{dS_{\rm cp}} h_{\nu \nu'} =& \frac{h_{\nu \nu'}}{\bar{p}_0} \left[ \omega_{\rm e}^{(\rm L)} \delta_{\nu,1} +\frac{1}{2} {\rm Tr} \, {\mathbf A}_0^{(\rm L)} - \omega_{\rm e}^{(\rm R)} \delta_{\nu',1} \right. \\
    \nonumber
    &\left.  - d-\frac{1}{2} {\rm Tr}\, {\mathbf A}_0^{(\rm R)} \right] \\
    \frac{d}{dS_{\rm cp}} h_{\nu \nu'} =& \frac{h_{\nu \nu'}}{\bar{p}_0} \left( E^{(\rm L)}- E^{(\rm R)}\right),
\label{der_h}
\end{align}
where $E^{(\rm{L} / \rm{R})}$ are the energies of the local L/R wavefunctions and ${\mathbf A}_{\perp}$ is the matrix
${\mathbf A}$ projected onto the space orthogonal to the MAP. From \eqn{der_h}, it is evident that the
TM element will only be invariant with respect to variation in $S_{\rm cp}$ if the two local L/R states are in
resonance. Otherwise, the TM element depends on the position of the connection point and the dependence is stronger
for a larger mismatch in the local energies.
 
The dependence on the connection point can be traced to the neglected integral in the derivation of Herring formula
in \eqn{herring_derivation} involving $\phi_{\nu}^{(\rm{L})}\phi_{\nu'}^{(\rm{R})}$ term.
Inclusion of this term, transforms the formula for the TM element into
\begin{align}
        \nonumber
        \tilde{h}_{\nu \nu'} =& \frac{1}{2} \int \left( \phi_{\nu'}^{(\rm R)} \frac{\partial}{\partial S} \phi_{\nu}^{(\rm L)} -
        \phi_{\nu}^{(\rm L)} \frac{\partial}{\partial S} \phi_{\nu'}^{(\rm R)} \right) \delta(f_{\rm D}(\mathbf{x})) d\mathbf{x} \\
        &+(E^{(\rm{L})}-E^{(\rm{R})})\int_{\rm{L}} \phi_{\nu}^{(\rm{L})}\phi_{\nu'}^{(\rm{R})} d\mathbf{x},
\label{herring_deriv_2}
\end{align}
where the second integral is taken over the space on the `left' side of the dividing plane. 
We then differentiate the extended expression, \eqn{herring_deriv_2},
with respect to $S_{\rm cp}$, and observe that the derivative of the additional term in \eqn{herring_deriv_2}
is the integral over the dividing plane. The derivative of the TM element becomes
\begin{align}
        \nonumber
        \frac{d}{d S_{\rm{cp}}} \tilde{h}_{\nu \nu'} = &(E^{(\rm{L})}-E^{(\rm{R})}) \frac{h_{\nu \nu'}}{\bar{p}_0} + \\
        &(E^{(\rm{L})}-E^{(\rm{R})}) \int \phi_{\nu}^{(\rm L)}\phi_{\nu'}^{(\rm R)} \delta(f_{\rm D}(\mathbf{x})) d\mathbf{x},
\label{herring_deriv_3}
\end{align}
where use has been made of \eqn{der_h}.
We note now that the surface integral in \eqn{herring_deriv_3} differs from the `old version'
of $h_{\nu \nu'}$ only by the momentum term $\bar{p}_0$. However, this factor is taken to be
constant on the dividing plane and it can be taken out of the integral, which leads to
\begin{equation}
\frac{d}{d S_{\rm{cp}}} \tilde{h}_{\nu \nu'} = (E^{(\rm{L})}-E^{(\rm{R})}) \frac{h_{\nu \nu'}}{\bar{p}_0} -
(E^{(\rm{L})}-E^{(\rm{R})}) \frac{h_{\nu \nu'}}{\bar{p}_0} = 0.
\end{equation}
While it would seem obvious to use the instanton wavefunctions in \eqn{localized_wavefunctions}
and compute the additional term that arises in \eqn{herring_deriv_3} in order to eliminate
the dependence on $S_{\rm cp}$, the semiclassical approximation breaks down beyond the barrier
region and would lead to a divergence of the integral.
We are thus left with no option but to neglect the term and treat it as a source of error
in the TM element.

We tried to estimate the size of the error term, introduced above, by calculating the overlap of
the harmonic oscillator wavefunctions centered at the two minima.
The overlap is obtained following the method of Ref.~\onlinecite{Berger1998},
where it was used to calculate Frank-Condon factors between the shifted harmonic oscillators.
The calculation requires the knowledge of molecular geometries and harmonic frequencies
in the two minima and does not add to the overall computational cost.
The errors for the partially deuterated malonaldehyde were found to be small and
are listed in the parenthesis in Table \ref{tm_pdm}.
We found that it was safe to neglect the error terms unless the connection point is
moved far from the position of the barrier top, or unless the energy difference
between the two states is large in relation to the barrier height.
However, we note that as the energy difference increases, the contribution of
the TM element to the energy level decreases, as seen from the second-order
perturbation treatment,
\begin{align}
    \nonumber
    E_i^{(\rm L), (2)} = \sum_j \frac{h_{ij}^2}{E_i^{(\rm L)}-E_j^{(\rm R)}}, \\
    E_i^{(\rm R), (2)} = \sum_j \frac{h_{ji}^2}{E_i^{(\rm R)}-E_j^{(\rm L)}}.
\end{align}
Since $h_{ij}$ are expected to be small in the deep tunneling regime,
the contributions of the states that lie far from the resonance
quickly approach zero. Thus, even for larger energy differences,
the computed energy levels remain stable.

Dependence of the TM elements on the position of
the dividing plane was also tested by moving the dividing plane
from $0.25 S_{\rm tot}$ to $0.75 S_{\rm tot}$, where $S_{\rm tot}$ is the MAP length.
All the matrix elements displayed monotonous dependence on $S_{\rm cp}$, with an inflection point
located near the position of the barrier top, and a region of relative stability around it.
As the dividing plane is moved away from the barrier top, some TM elements begin to change
significantly. These TM elements connect the states with a significant disparity in their local energies.
For example, the TM elements between the $(\nu_2+\nu_2)$ state and the GS or the $\nu_2$-excited state
change from $-11 \ \rm{cm}^{-1}$ to $-55 \ \rm{cm}^{-1}$ and from $-36 \ \rm{cm}^{-1}$ to $-80 \ \rm{cm}^{-1}$,
respectively. However, a large energy difference between the involved states also implies
that the TM elements do not profoundly affect the overall energy.
The GS TS varies by $\approx 7 \ \rm{cm}^{-1}$ in this region. If we limit ourselves to
the region between $0.4 S_{\rm tot}$ and $0.6 S_{\rm tot}$, the variation of the GS TS is
only $1.5 \ \rm{cm}^{-1}$, which is  within the error of the VCI method used to compute
the local energies. Similar errors are observed for the other energy levels.
The method thus gives best results when the dividing plane is placed in the vicinity
of the barrier top.

\section{Vibrational self consistent field (VSCF) and vibrational configuration interaction (VCI)}
The idea behind the VSCF approach is to approximate the vibrational wavefunction by a Hartree product of
single-mode (1M) functions as
\begin{equation}
    \psi(\mathbf q) = \phi_0^{(1)}(q_1) \dots \phi_0^{(N)}(q_N),
\label{VSCF_wavefunction}
\end{equation}
where $q_i$ is the $i$-th normal mode coordinate. The above form can efficiently be employed in combination
with the $n$-mode representation \cite{Bowman2003,Rauhut2004} of the potential
\begin{equation}
    V(\mathbf q) = V_{\rm min}+\sum_{i=1}^{N} V_i^{(\rm 1M)}(q_i)+\sum_{i=1}^{N-1} \sum_{j=i}^{N} V_{ij}^{(\rm 2M)}(q_i, q_j) + \dots
\label{VSCF_potential}
\end{equation}
In this paper, the expansion was truncated at the two-mode (2M) terms. Using wavefunction in
\eqn{VSCF_wavefunction} and variational principle, it can be shown \cite{Carter1998,Christoffel1982,Bowman1979}
that the optimal 1M functions satisfy a set of coupled 1M equations
\begin{align}
    \nonumber
    &\frac{1}{2}\frac{d^2}{dq_i^2}\phi^{(i)}(q_i)+( V_{\rm min}+ V_i^{(\rm 1M)}(q_i) + \\
    &\sum_{\substack{
          j=1  \\ j\ne i}}^{N_{\rm dof}} \bra{\phi_0^{(j)}(q_j)} V_{ij}^{(\rm 2M)}(q_i, q_j) \ket{\phi_0^{(j)}(q_j)} ) \phi^{(i)}(q_i) = \varepsilon_i\phi^{(i)}(q_i),
\label{VSCF_equations}
\end{align}
which can be solved via an SCF algorithm. The 1M functions $\phi_0^{i}$ are expanded in
a basis set $\{ \chi_j^{(i)} \}$. In this paper, the lowest $N_{\rm basis}$ states of
the harmonic oscillator (HO) are used. The 1M and 2M potential energy terms in \eqn{VSCF_potential}
are fitted to polynomials of order $N_{\rm fit}$ as
\begin{align}
    \nonumber
    &V_i^{(\rm 1M)}(q_i)=\sum_{j=2}^{N_{\rm fit}} C^{(i)}_j q_i^j \\
    &V_{ij}^{(\rm 2M)}(q_i, q_j)=\sum_{k=1}^{N_{\rm fit}-1}\sum_{l=1}^{N_{\rm fit}-k} C^{(ij)}_{kl} q_i^kq_j^l.
\end{align}
These choices enable us to compute the potential matrix elements exactly. For that purpose, we calculate
the matrices $\mathbf{Q}_i^{(j)}$, that represent operators $q_i^j$ in the HO basis.
This can be done recursively using ladder operators
\begin{align}
    \nonumber
& \bra{\chi_j^{(i)}} q_i^n \ket{\chi_k^{(i)}} = 
    \frac{1}{\sqrt{2\omega_i}}\bra{\chi_j^{(i)}} q_i^{n-1}(\hat{a}_i+\hat{a}^{\dag}_i) \ket{\chi_k^{(i)}} = \\
& = \sqrt{\frac{k}{2\omega_i}} \bra{\chi_j^{(i)}} q_i^{n-1} \ket{\chi_{k-1}^{(i)}}+ 
    \sqrt{\frac{k+1}{2\omega_i}} \bra{\chi_j^{(i)}} q_i^{n-1} \ket{\chi_{k+1}^{(i)}}.
\label{VSCF_recursion}
\end{align}
The matrices $\mathbf{Q}_i^{(j)}$ are then stored and the SCF algorithm is started using the initial guess
$\phi_0^{(i)}=\chi_0^{(i)}$. The effective Hamiltonian in \eqn{VSCF_equations} is constructed for each mode
in the HO basis, diagonalized and the obtained 1M functions corresponding to the lowest eigenvalue taken
as new $\phi_0^{(i)}$. Once VSCF has converged, the obtained virtual 1M functions are
used to construct the VCI Hamiltonian
\begin{equation}
H_{IJ}=\bra{\phi_{i_1}^{(1)} \dots \phi_{i_N}^{(N)}} \hat{H} \ket{\phi_{j_1}^{(1)} \dots \phi_{j_N}^{(N)}}.
\end{equation}
The matrix elements of $H_{IJ}$ are computed using $\mathbf{Q}_i^{(j)}$ matrices and the coefficients
of 1M functions in the HO basis. Its eigenvalues represent VCI energies, while the eigenvectors can be
used to determine the dominant configuration $\phi_{i_1}^{(1)} \dots \phi_{i_N}^{(N)}$ and tell us
which normal modes are excited in the particular state.

\begin{figure} [htbp]
\begin{center}
\rotatebox{0}{ \resizebox{8cm}{!}
{\includegraphics[width=8cm]{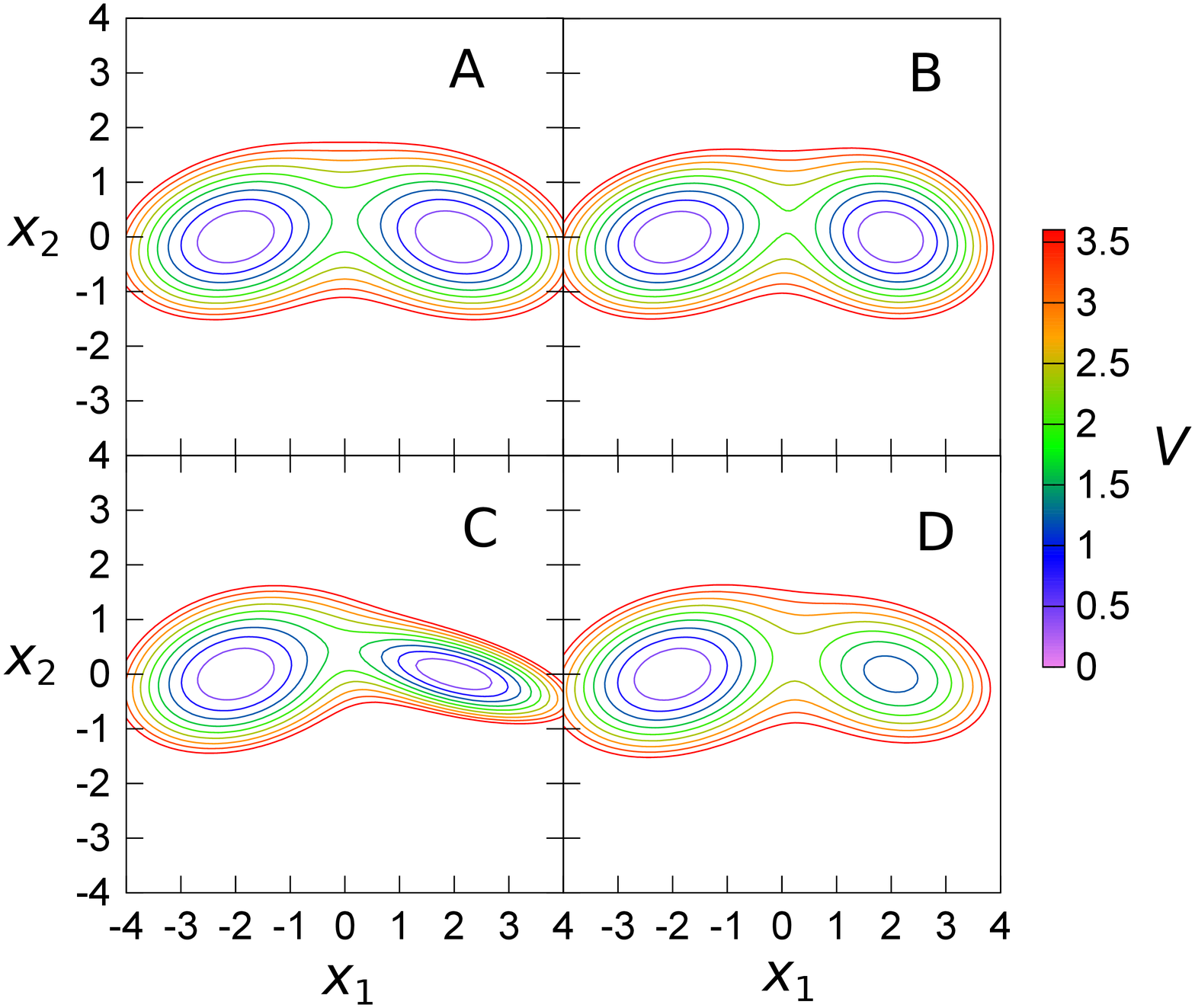}}}
\caption{Potential energy surface of the 2D model given by \eqn{sym_pot}.
Top left panel corresponds to the symmetric potential, top right to $\omega_1^{\rm (R)} > \omega_1^{\rm (L)}$,
bottom left to $\omega_2^{\rm (R)} > \omega_2^{\rm (R)}$, and bottom right to $d>0$, with other parameters set
equal to the symmetric case.}
\label{fig_1}
\end{center}
\end{figure}

\begin{figure*} [htbp]
\begin{center}
\rotatebox{0}{ \resizebox{16cm}{!}
{\includegraphics[width=16cm]{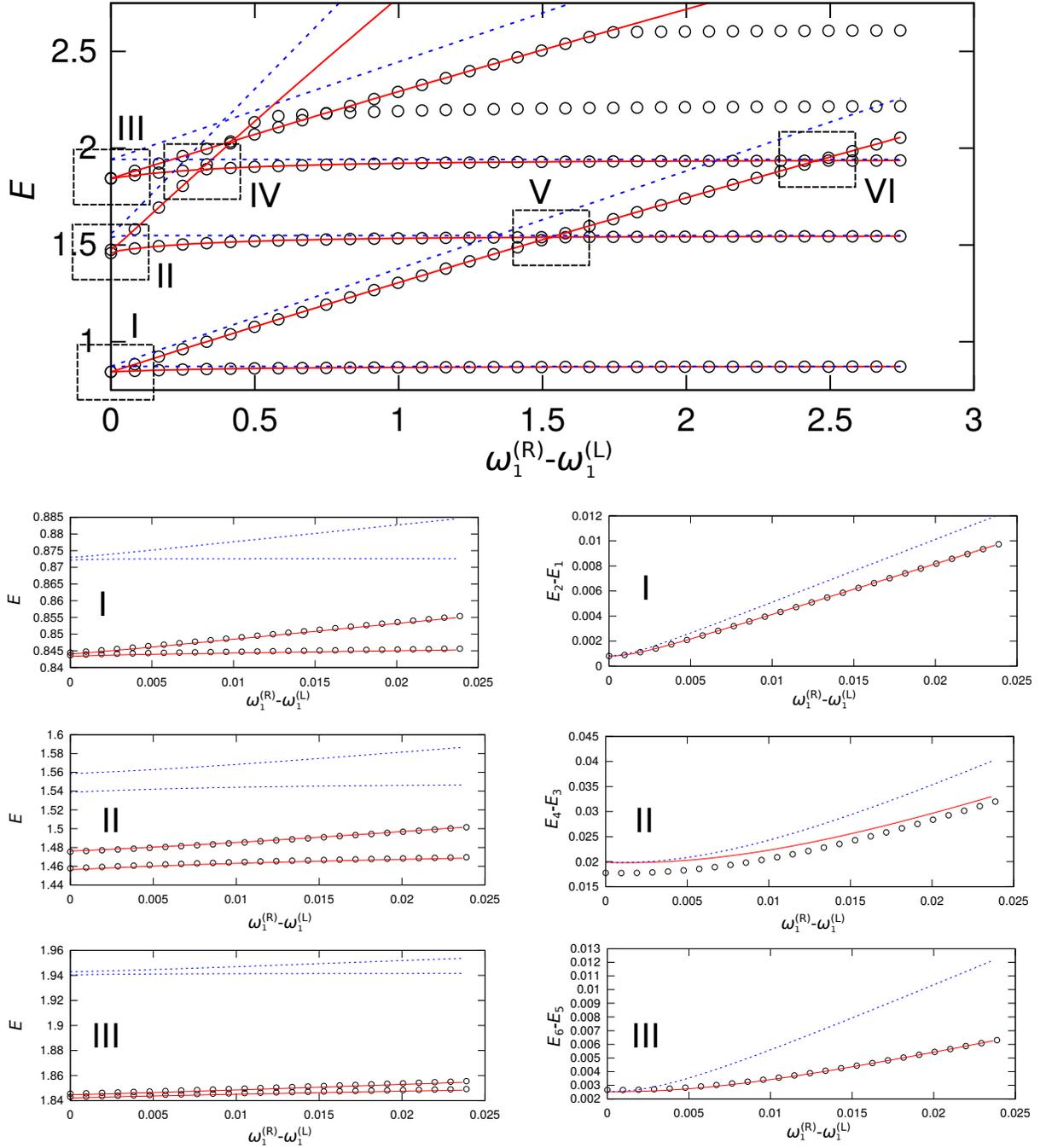}}}
\caption{Dependence of vibrational energies of the lowest 6 states in the double-well potential given
by \eqn{sym_pot} on $\omega_1^{(\rm R)}$. Circles represent quantum-mechanical values, blue lines are obtained 
using instanton method with harmonic energies, red lines are obtained using a combined VCI/instanton approach.
Frames I-III in the top panel are shown magnified in the left column panels below, and the dependence of
the associated tunneling splittings on $\omega_1^{(\rm R)}$ in the right column panels below.}
\label{fig_2}
\end{center}
\end{figure*}

\begin{figure} [htbp]
\begin{center}
\rotatebox{0}{ \resizebox{8cm}{!}
{\includegraphics[width=8cm]{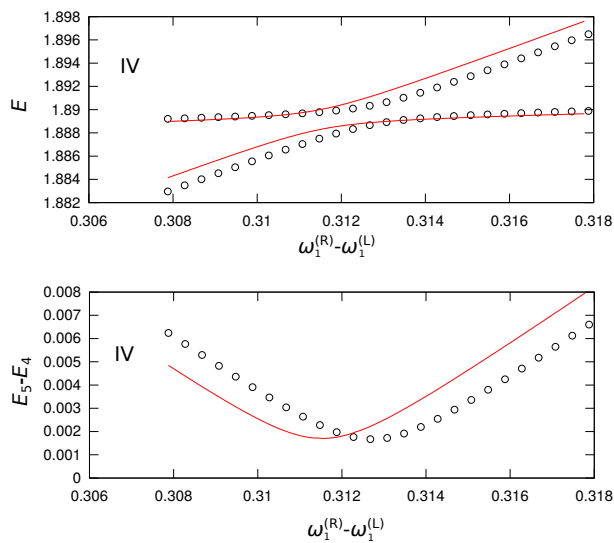}}}
\caption{Dependence of vibrational energies and tunneling splittings in the 2D model potential given by
\eqn{sym_pot} on the frequency $\omega_1^{(\rm R)}$ in the region of the avoided crossing
between the first ($\omega_1^{\rm (R)}$-) excited state in
the right minimum and the second ($\omega_2^{\rm (L)}$-) excited state in the left minimum, shown in frame
IV in the top panel of Figure \ref{fig_2}.
Circles represent quantum-mechanical values, while red lines represent values obtained using a combined
VCI/instanton approach.}
\label{fig_3}
\end{center}
\end{figure}

\begin{figure*} [htbp]
\begin{center}
\rotatebox{0}{ \resizebox{16cm}{!}
{\includegraphics[width=16cm]{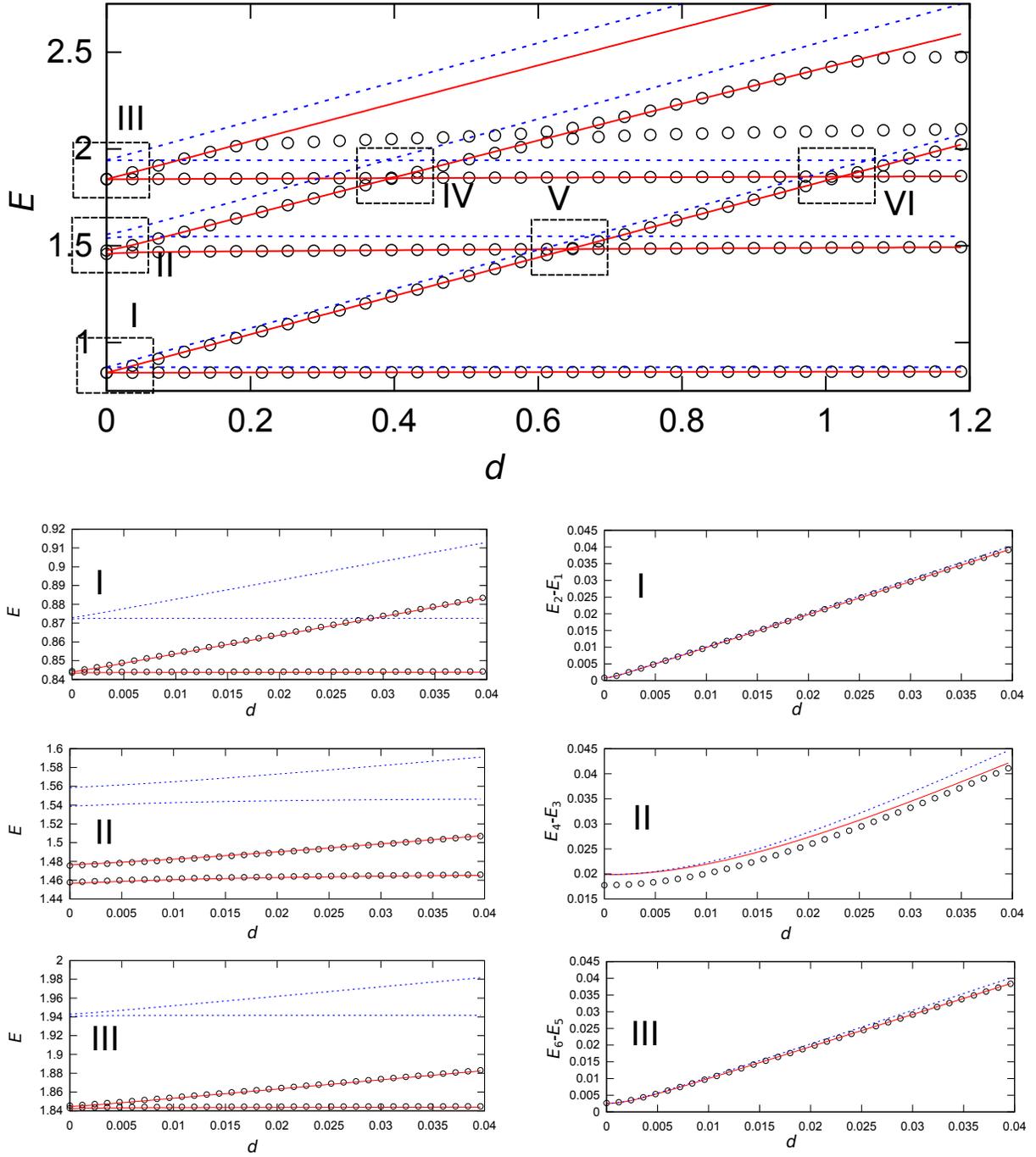}}}
\caption{Dependence of vibrational energies of the lowest 6 states in the double-well potential given
by \eqn{sym_pot} on the energy shift $d$ of the right well. Circles represent quantum-mechanical values,
blue lines are obtained  using instanton method with harmonic energies, red lines are obtained using
a combined VCI/instanton approach. Frames I-III in the top panel are shown magnified in the left column
panels below, and the dependence of the associated tunneling splittings on $d$ in the right column panels
below.}
\label{fig_4}
\end{center}
\end{figure*}

\begin{figure} [htbp]
\begin{center}
\rotatebox{0}{ \resizebox{8cm}{!}
{\includegraphics[width=8cm]{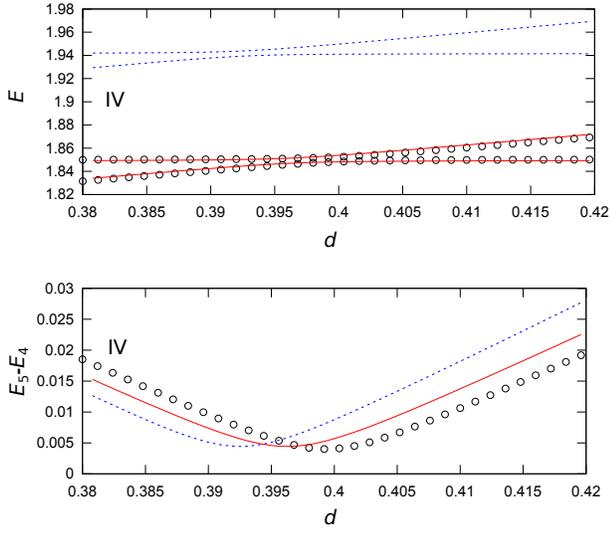}}}
\caption{Dependence of vibrational energies and tunneling splittings in the 2D model potential given by
\eqn{sym_pot} on the energy shift $d$ of the right well in the region of the avoided crossing
between the first ($\omega_1^{\rm (R)}$-) excited state in the right minimum and
the second ($\omega_2^{\rm (L)}$-) excited state in the left minimum, shown in frame
IV in the top panel of Figure \ref{fig_4}.
Circles represent quantum-mechanical values, blue lines are obtained using instanton method with
harmonic energies, while red lines represent values obtained using a combined VCI/instanton approach.}
\label{fig_5}
\end{center}
\end{figure}

\begin{figure*} [htbp]
\begin{center}
\rotatebox{0}{ \resizebox{16cm}{!}
{\includegraphics[width=16cm]{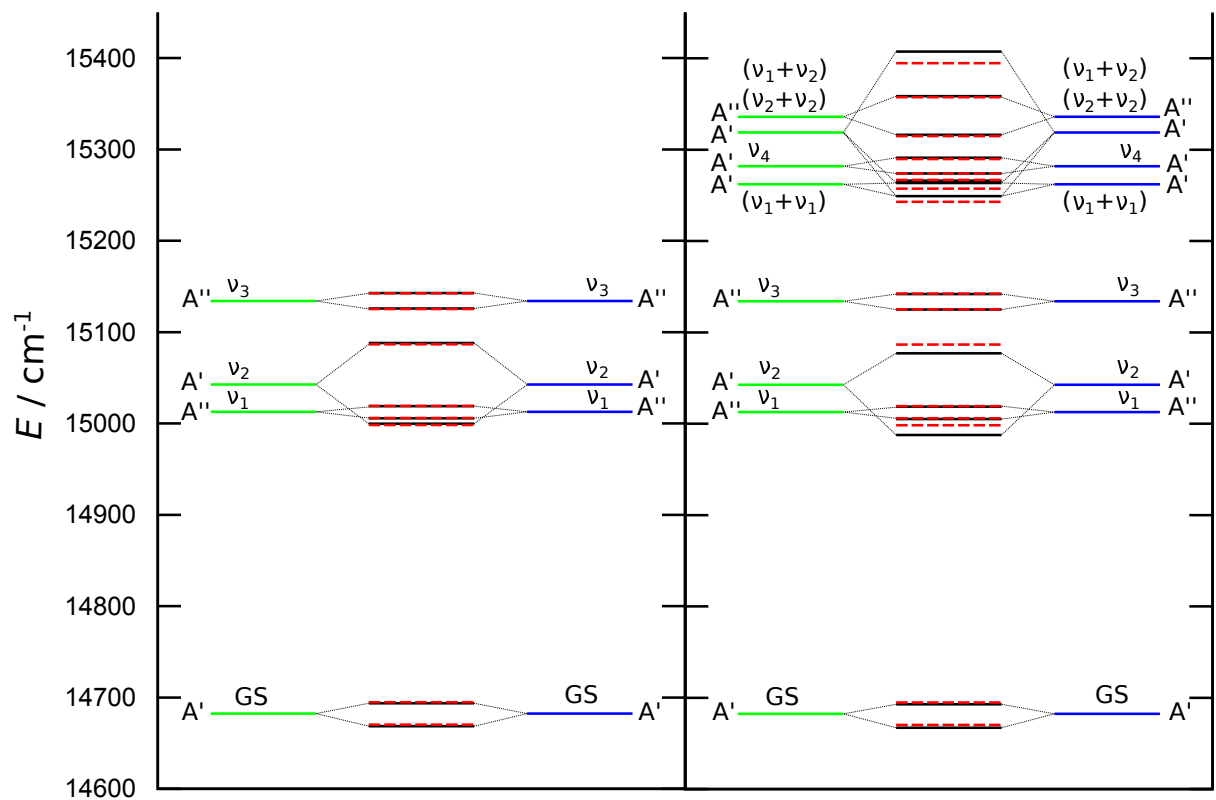}}}
\caption{Vibrational tunneling spectrum of the lowest 8 (left panel) and 16 (right panel) states of malonaldehyde.
Green and blue lines represent VCI energies of local wavefunctions in the D7 and D9 minima, respectively.
Dashed red lines are obtained using a $2 \times 2$ tunneling matrix (TM) model. Black lines in the left panel
represent energies from an $8 \times 8$ TM model, and in the right panel, they represent energies from
a $16 \times 16$ model. See text for details.}
\label{fig_6}
\end{center}
\end{figure*}

\begin{figure*} [htbp]
\begin{center}
\rotatebox{0}{ \resizebox{16cm}{!}
{\includegraphics[width=16cm]{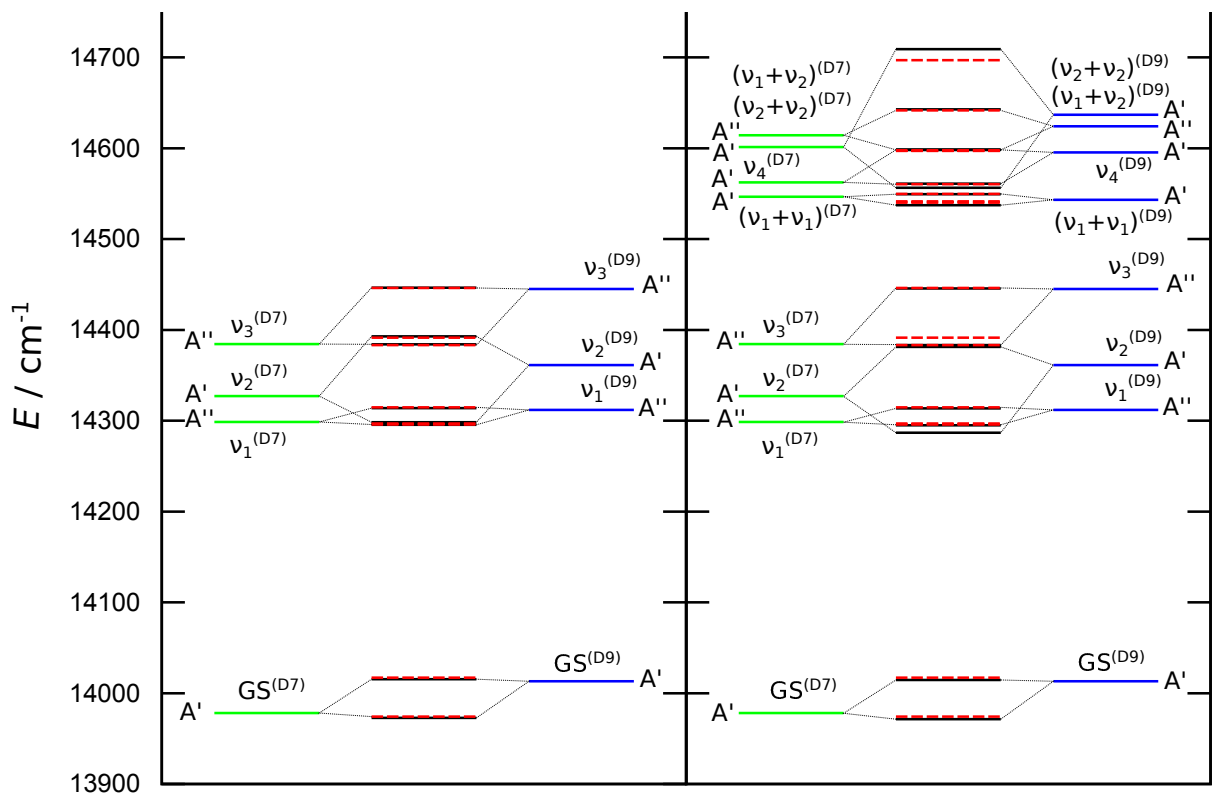}}}
\caption{Vibrational tunneling spectrum of the lowest 8 (left panel) and 16 (right panel) states of
partially deuterated malonaldehyde.
Green and blue lines represent VCI energies of local wavefunctions in the D7 and D9 minima, respectively.
Dashed red lines are obtained using a $2 \times 2$ tunneling matrix (TM) model. Black lines in the left panel
represent energies from an $8 \times 8$ TM model, and in the right panel, they represent energies from
a $16 \times 16$ model. See text for details.}
\label{fig_7}
\end{center}
\end{figure*}

\bibliography{references,new_references}

\end{document}